\documentclass[aps,prx,twocolumn,superscriptaddress,noeprint,longbibliography, nofootinbib,twocolumn]{revtex4-1}

\usepackage{graphicx}
\usepackage{bm}
\usepackage{physics}
\usepackage{xcolor}
\usepackage{enumitem}
\usepackage{amsmath, amssymb}
\usepackage[normalem]{ulem}
\usepackage{natbib}
\usepackage{comment}

\newcommand{\be}{\begin{equation} }
\newcommand{\ee}{\end{equation} }
\newcommand{\ba}{\begin{eqnarray} }
\newcommand{\ea}{\end{eqnarray} }

\newcommand{\bit}{\begin{itemize}}
\newcommand{\eit}{\end{itemize}}

\graphicspath{{Figures/}}

\usepackage[colorlinks=true]{hyperref}
\hypersetup{citecolor = blue}

\begin{document}

\title{ Photon-ALP beam propagation from Mrk 501}

\author{L. J. Dong}
\affiliation{Department of Physics, Yunnan Normal University, Kunming, Yunnan, 650092, China}
\author{Y. G. Zheng}
\thanks{Corresponding author: Y.G. Zheng, \\ ynzyg@ynu.edu.cn}
\affiliation{Department of Physics, Yunnan Normal University, Kunming, Yunnan, 650092, China}
\affiliation{Key Laboratory of Colleges and Universities in Yunnan Province for High-energy Astrophysics, Kunming, Yunnan, 650500, China}
\author{S. J. Kang}
\affiliation{School of Physics and Electrical Engineering, Liupanshui Normal University, Liupanshui, Guizhou, 553004, China}
\author{C. Y. Yang}
\affiliation{Yunnan Observatories, Chinese Academy of Sciences, Kunming 650011, China}
\affiliation{Key Laboratory of Astroparticle Physics of Yunnan Province, Kunming 650091, China}

\begin{abstract}
The very high energy (VHE, E $>$ $100 \mathrm~{GeV}$) $\gamma$-ray observations offer a possibility of indirectly detecting the presence of axion-like particles (ALPs). The paper focuses on detecting photon-ALP oscillations on $\gamma$-ray spectra from distant sources in astrophysical magnetic fields. Strong evidence indicates that: (1) the photon-ALP oscillations can effectively decrease the photon absorption at energies of several tens of TeV -- caused by the extragalactic background light (EBL) -- to a level able to explain better the observational data; (2) the impact of magnetic-field models in photon-ALP beams crossing several magnetized media is significant. We revisit the expected signature for the photon-ALP oscillation effects on $\gamma-\gamma $ absorption in the TeV spectra of Mrk 501. The result issues that the photon-ALP beam propagation with mass $\mathrm{m_a}\sim10^{-10} eV$ and two-photon coupling constant $\begin{aligned}g_{a\gamma}\sim0.417\times10^{-11}GeV^{-1}\end{aligned}$ crossing reasonable magnetic field scenarios considered here can roughly reproduce the observed TeV $\gamma$-ray spectra for Mrk 501.
\end{abstract}

\maketitle

\section{Introduction}

The evidence has shown that the main elements of the universe are dark matter and dark energy \citep{2000PhST...85..210T}. Axions are linked to the global Peccei–Quinn symmetry introduced as a solution to the strong CP problem \citep{1977PhRvL..38.1440P,1978PhRvL..40..223W,1978PhRvL..40..279W}, and they are promising dark-matter candidate \citep{2011PhRvD..83l3526M,MARSH20161}. Moreover, axion-like particles (ALPs) are a generalization of the axion, and are very light spin-zero bosons predicted by a variety of extensions of the particle standard model \citep{1983PhLB..120..127P,2010PhRvD..81l3530A}. One of the most important phenomenological properties of the ALP is its two-photon coupling, which allows for photon-ALP mixing in the presence of external magnetic fields \citep{1978PhRvL..40..279W,2010AIPC.1223..128R,2011PhRvD..84j5030D}, leading to two distinctly different effects \citep{1988PhRvD..37.1356R}. One is that photon-ALP oscillations occur in the presence of an external magnetic field, and the $a\gamma\gamma $ coupling leads to interaction eigenstates that differ from propagation eigenstates, which is very similar to neutrino oscillations involving different masses. The other is the polarization change of photons crossing the magnetic field \citep{1988PhRvD..37.1356R,2023PhRvD.107d3006G,2023PhRvD.107j3007G,2023PhRvD.108h3017G,2024Univ...10..312G}. Generally speaking, this mixing effect can be exploited to search for ALPs in light-shining-through-the-wall experiments \citep{2010PhLB..689..149E,2008PhRvL.100h0402C}.

Photon-ALP oscillations provide an interesting way to significantly reduce the absorption of very-high-energy (VHE) photons by the pair production process ($\gamma_{\mathrm{VHE}}+\gamma_{\mathrm{EBL}}\rightarrow e^{+}+e^{-}$) at the extragalactic background light (EBL) above approximately 100 GeV. In this sense, numerous studies have been conducted to explore indications of photon-ALP oscillations in $\gamma$-ray spectra \citep{2012PhRvD..86g5024H, Meyer_2014,10.1093/mnras/stz1144, Zhou_2021,2023PhRvL.131y1001G}.
Two complementary realizations have been proposed: one involves the photon-ALP oscillation occurring in the turbulent extragalactic magnetic fields $B\sim0.1-1\mathrm~{nG}$ \citep{2007PhRvD..76l1301D,2011PhRvD..84j5030D}. However, there are no conclusive observations that the intergalactic magnetic field strength is close to the upper limit. The other is the presence of a $\gamma\rightarrow a$ conversion inside the blazar and the $a\rightarrow\gamma$ conversion in the Milky Way
\citep{2008PhRvD..77f3001S}. The magnetic field in blazars is highly complex and poorly understood. As a result, there is considerable uncertainty regarding whether the $\gamma\rightarrow a$ conversion occurs in these blazars.

In this paper, we discuss the photon-ALP oscillation effect which reduces the EBL absorption increasing the observed TeV spectrum from Markarian 501 (Mrk 501), while calculating the final photon survival probability $P_{\gamma\gamma}$ and the corresponding $\gamma$-ray spectral energy distribution (SED). We consider the case where photons are generated and oscillate into ALPs inside the jet with helical and tangled magnetic field and the alternative case of photon production in the jet with the simple toroidal magnetic field. Moreover, the photon-ALP beam propagates inside the Gaussian turbulent or structured cavity galaxy cluster magnetic field. Furthermore, photon propagation considers EBL absorption, while photon-ALP oscillation effects are negligible in extragalactic space. Finally, we add photon-ALP oscillation effects in the cases of Jansson $\&$ Farrar model or Pshirkov model in the Milky Way. We assume the $\gamma \rightarrow a$ conversion inside the jet and galaxy cluster and the $a\rightarrow\gamma$ conversion in the Milky Way. For Mrk 501 with the photon-ALP beam crossing reasonable magnetic scenarios and given a sufficiently small ALP mass, photon-ALP oscillation effects could leave an imprint on the $\gamma$-ray spectra.

In summary, the $\gamma$-ray flux is attenuated by interactions with the EBL, but photon-ALP conversions in the crossed magnetic fields can lead to an increase in the observed flux. We expect to find a signature of photon-ALP oscillation effects on the $\gamma$-ray spectra from Mrk 501. The paper is structured as follows. In Sec.\ref{sec:dataset}, we introduce photon-ALP oscillation effects. In Sec.\ref{sec:freezeout}, we describe the propagation of the photon-ALP beam in different magnetic fields, whereas in Sec.\ref{sec:Magnuss} we describe the process of spectral analysis of the blazar and apply our model to predict the expected $\gamma$-ray flux of Mrk 501. In Sec.\ref{sec:Conclusion and Discussion}, we present our discussion and conclusion. Throughout the paper, we use the natural units $\hbar=c=k=1$.

\section{Photon-ALP oscillations} \label{sec:dataset}

The effective Lagrangian for the oscillation between photons and ALPs is expressed as \citep{1988PhRvD..37.1237R}
\begin{equation}
\mathcal{L}=\mathcal{L}_{a\gamma}+\mathcal{L}_{\mathrm{EH}}+\mathcal{L}_{a},
\label{Eq:1}
\end{equation}
where $\mathcal{L}_{a\gamma}=-\frac{1}{4}g_{a\gamma}F_{\mu\nu}\tilde{F}^{\mu\nu}a$ corresponds to the photon-ALP mixing, $F_{\mu\nu}$ is the electromagnetic field strength, $\tilde{F}^{\mu\nu}$ its dual,  and $a$ is the ALP field strength. $\mathcal{L}_{\mathrm{EH}}$ represents the effective Euler-Heisenberg Lagrangian \citep{1984PhT....37i..80I} and $\mathcal{L}_{a}=\frac{1}{2}\partial_{\mu}a\partial^{\mu}a-\frac{1}{2}m_{a}^{2}a^{2}$ denotes the ALP mass kinetic terms.

We consider monochromatic, unpolarized photon-ALP beams with energy $\text{E}$ and wave vector $\mathbf{k}$ propagating through a cold, magnetized, and ionized medium, as described in \cite{2011PhRvD..84j5030D}.
The external magnetic field $\mathbf{B}$ is assumed to be uniform, and the electron number density is described by $n_{e}$. We use an orthogonal reference frame with the $\text{z}$ axis aligned along $\mathbf{k}$, and the $\text{x}$ and $\text{y}$ axes are chosen arbitrarily \citep{2011PhRvD..84j5030D}. The equations of motion for the photon beam propagation are expressed in a $\text{Schr$\ddot{o}$dinger-like}$ form \citep{1988PhRvD..37.1237R}
\begin{equation}
\left(i\frac{\mathrm{d}}{\mathrm{d}z}+E+\mathcal{M}_0\right)\Psi(z)=0,
\end{equation}
where $\Psi(z)=(A_{x}(z),A_{y}(z),a(z))^{T}$, $A_{x}(z),A_{y}(z)$ represents the polarisation states along $x$ and $y$, respectively. The mixing matrix $\mathcal{M}_{0}$, which neglects Faraday rotation effects, is given by
\begin{equation}
\mathcal{M}_0=\begin{pmatrix}\Delta_\perp&0&0\\0&\Delta_{||}&\Delta_{a\gamma}\\0&\Delta_{a\gamma}&\Delta_a\end{pmatrix},
\end{equation}
where $\Delta_{\perp}=\Delta_{\mathrm{pl}}+2\Delta_{\mathrm{QED}}+\Delta_{\mathrm{CMB}}$ 
represents the effect of photon propagation in a plasma, where $\Delta_{\mathrm{pl}}\equiv-\omega_{\mathrm{pl}}^{2}/2E$, $\Delta_{\mathrm{QED}}=\alpha E/(45\pi)(B/(B_{\mathrm{cr}}))^{2}$ and $\Delta_{\mathrm{CMB}}$ account for the plasma term, the QED vacuum polarization and photon-photon dispersion in the cosmic microwave background (CMB) term, respectively. $\alpha$ represent the fine-structure constant and $B_{\mathrm{cr}}$ is the critical magnetic field. $\Delta_{||}=\Delta_{\mathrm{pl}}+7/2\Delta_{\mathrm{QED}}+\Delta_{\mathrm{CMB}}$  corresponds to the QED vacuum polarization. $\Delta_{a}=-m_{a}^{2}/(2E)$ is the ALP mass term. $\Delta_{a\gamma}=g_{a\gamma}B/2$ denotes the photon-ALP mixing term \citep{1988PhRvD..37.1237R}. More suitable numerical values for the different $\Delta$ terms are suggested in \cite{2012PhRvD..86g5024H}.

We examine the propagation of the photon-ALP beam from its origin in the BL Lac jet to Earth, aiming to estimate the photon survival probability. Given that the general case involves the magnetic field B not being aligned with the y-axis but forming an arbitrary angle $\psi$ with it, the z-axis is defined as the propagation direction. $N_{d}$ represents the total number of domains crossed by the beam from the blazar to earth. The code divides the calculation into $N_{d}$ steps along the z-axis, assuming a constant magnetic field for each step length dz. The transfer matrix describes the propagation of the photon-ALP beam for a fixed arbitrary choice of angles $\{\psi_{n}\}_{1\leq n\leq N_{d}}$ is \citep{2011PhRvD..84j5030D, 2010ApJ...712..238F}
\begin{equation}
\mathcal{U}(z_{N_{d}},z_1;\psi_{N_{d}},\ldots,\psi_1;E)=\prod_{i=1}^{N_{d}}\mathcal{U}(z_{i+1},z_i;\psi_i;E).
\label{eq:4}
\end{equation}

After the mixing in $N_{d}$ consecutive steps, the photon survival probability is given by
\begin{equation}
\begin{split}
P_{\gamma\gamma}^{\rm ALP} &= \mathrm{Tr}\left((\rho_{1}+\rho_{2})\mathcal{U}(z_{N_{d}},z_1;\psi_{N_{d}},\ldots,\psi_1;E)\right. \\
&\quad \left.\rho(0)\mathcal{U}^\dagger(z_{N_{d}},z_1;\psi_{N_{d}},\ldots,\psi_1;E)\right),
\end{split}
\label{eq:5}
\end{equation}
where $\rho(0)=(1/2){diag}(1,1,0)$, $\rho_1={diag}(1,0,0)$, $\rho_2={diag}(0,1,0)$. Details of the model are described in this webpage \footnote{ https://gammaalps.readthedocs.io}.

\section{ALP scenario} \label{sec:freezeout}

We aim to study a complete propagation scheme where the VHE photon-ALP beam originates from the jet and galaxy cluster, up to the Milky Way, while the magnetic field is ubiquitous in this propagation process. Photon-ALP conversions take place within the jet magnetic fields of blazars, where a significant amount of ALPs can be produced \citep{2015PhLB..744..375T, 2021PhRvD.103b3008D}, which explains why flat-spectrum radio quasars (FSRQs) with energies exceeding 30 $GeV$ can be observed \citep{2012PhRvD..86h5036T}. Additionally, photon-ALP oscillations can occur in the magnetic fields of galaxy clusters, leading to irregularities in the observed spectra \citep{2016PhRvL.116p1101A, Abdalla_2021}. Furthermore, the photon-ALP oscillation effect effectively suppresses photon attenuation by preventing their interaction with the Extragalactic Background Light (EBL) in intergalactic space. \citep{2011PhRvD..84j5030D, 2007PhRvD..76l1301D}. Finally, the observed spectra of blazars exhibit alterations when different models of the Milky Way are considered. \citep{2008PhRvD..77f3001S}.  Actually, the photon-ALP oscillation effect from the magnetic fields of the host galaxy is neglected \citep{2012PhRvD..86h5036T}.

\begin{figure*}
\centering
 \includegraphics[height=6cm,width=8.8cm]{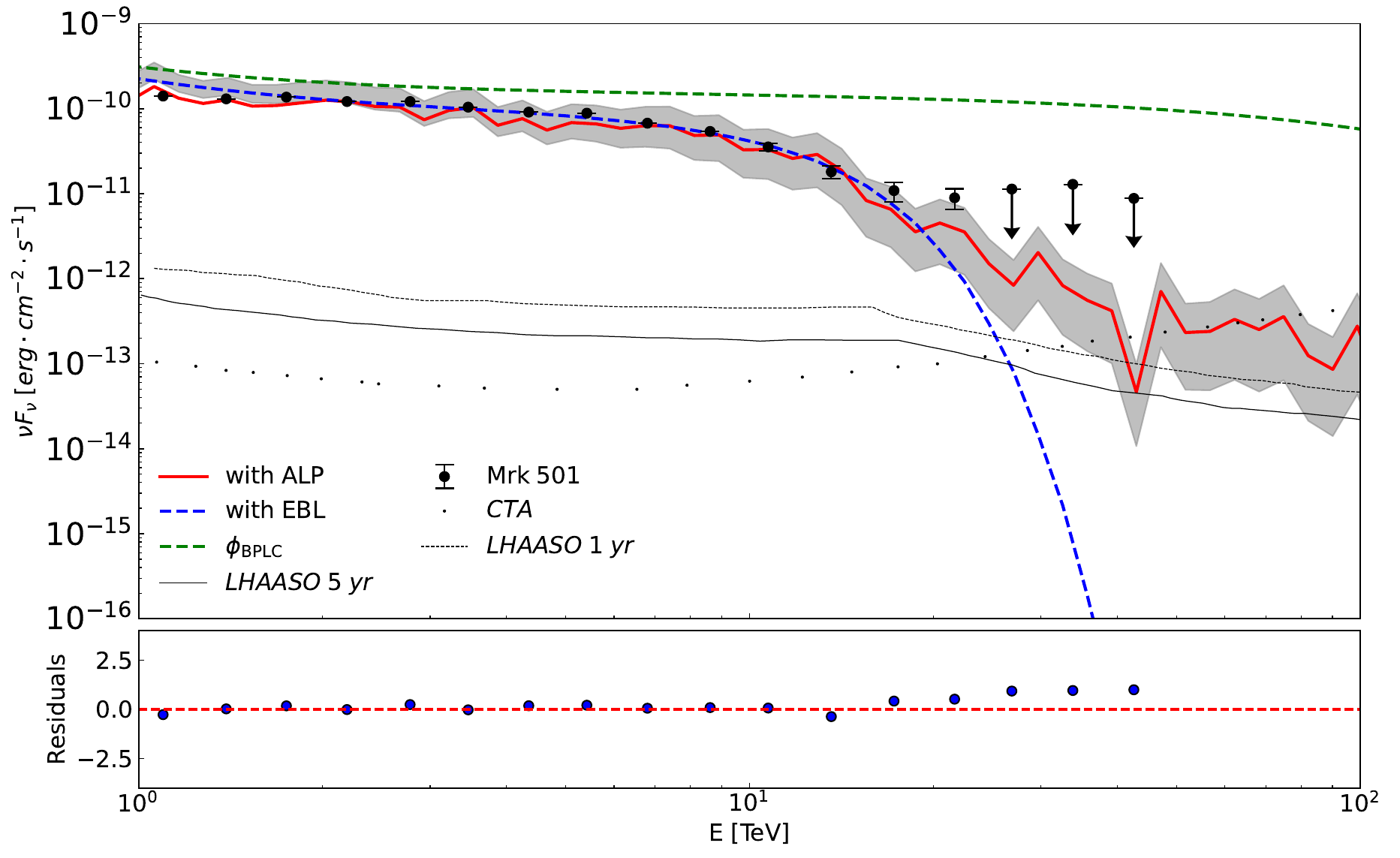}
 \includegraphics[height=6cm,width=8.8cm]{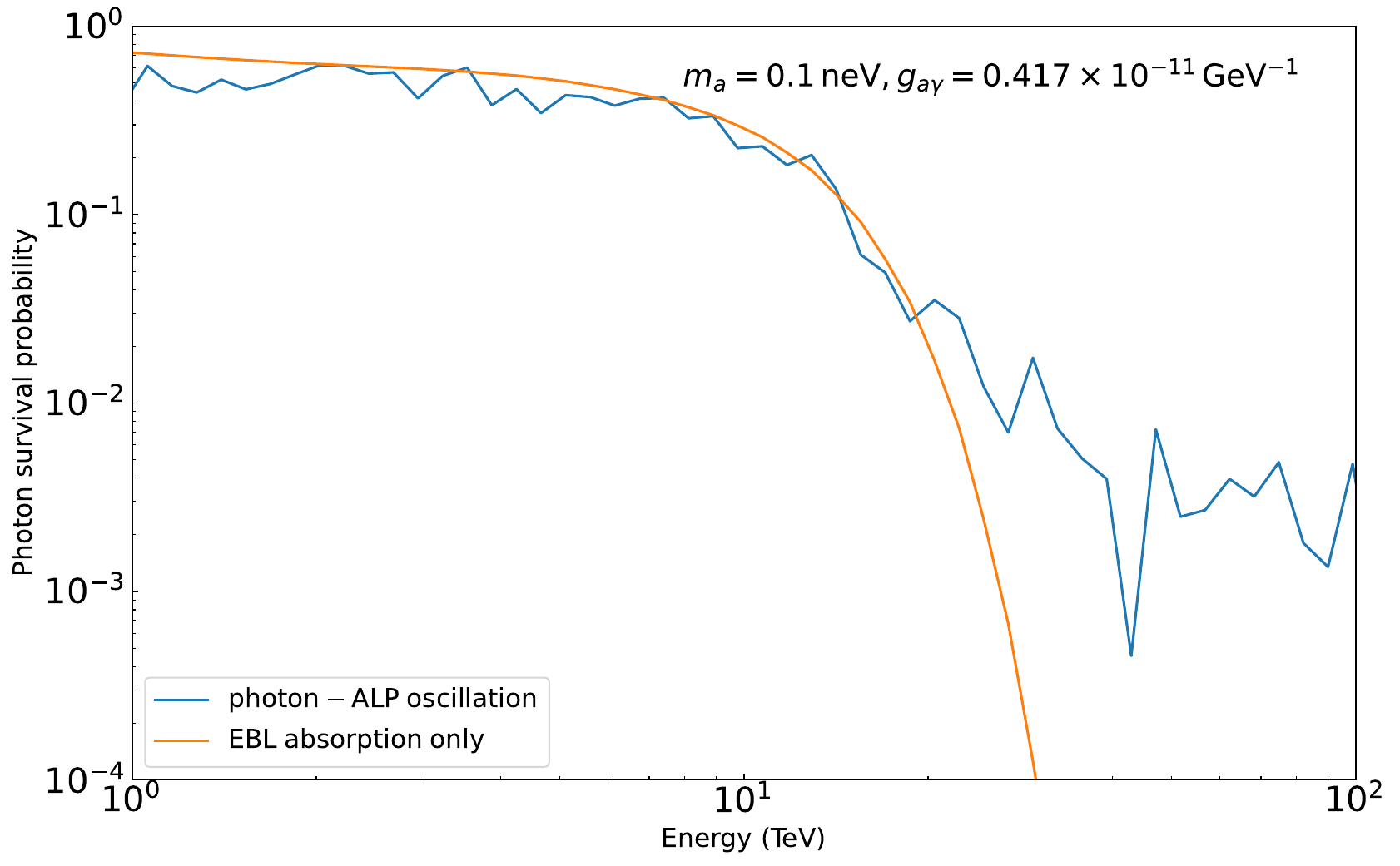}
 \includegraphics[height=6cm,width=8.8cm]{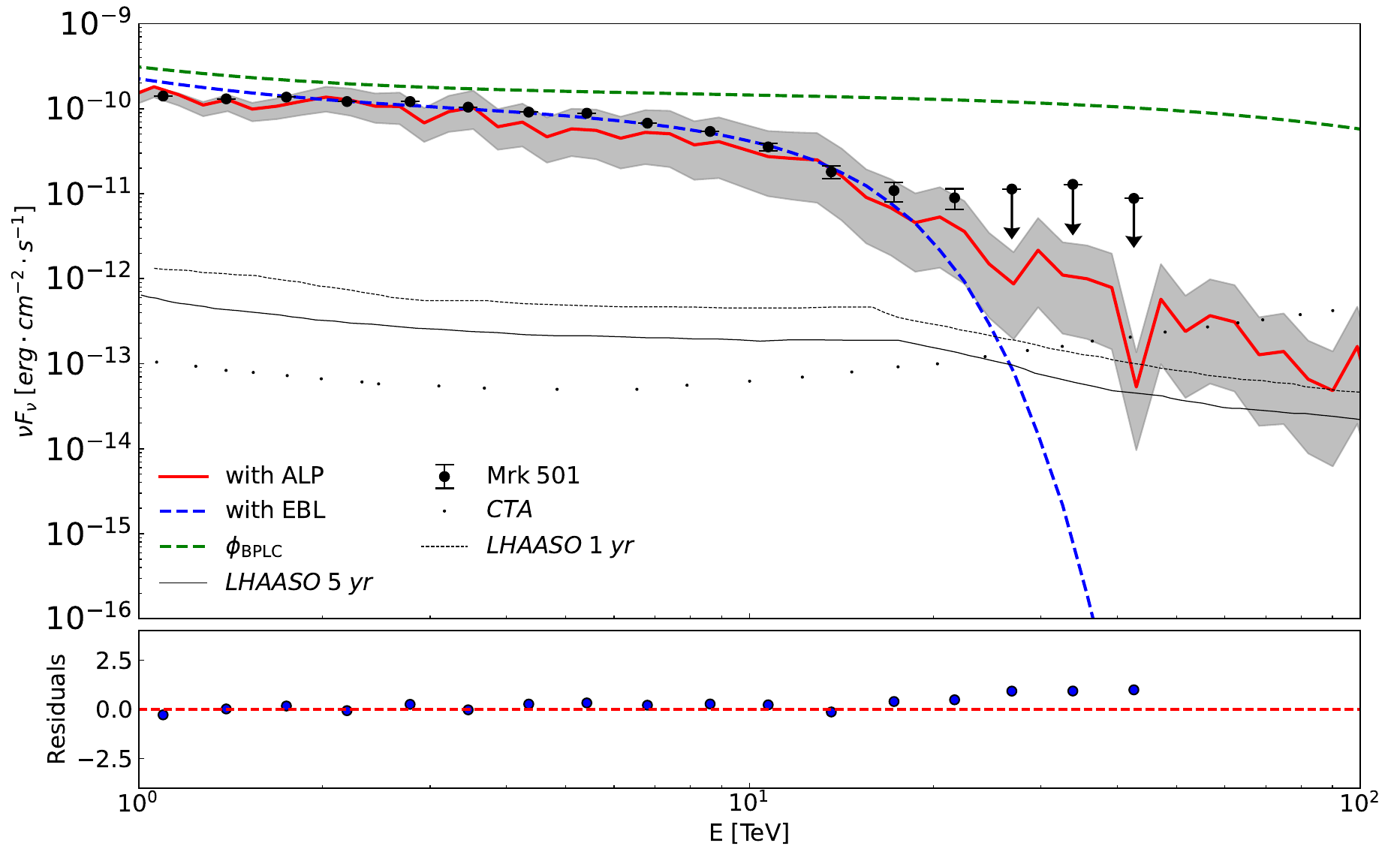}
 \includegraphics[height=6cm,width=8.8cm]{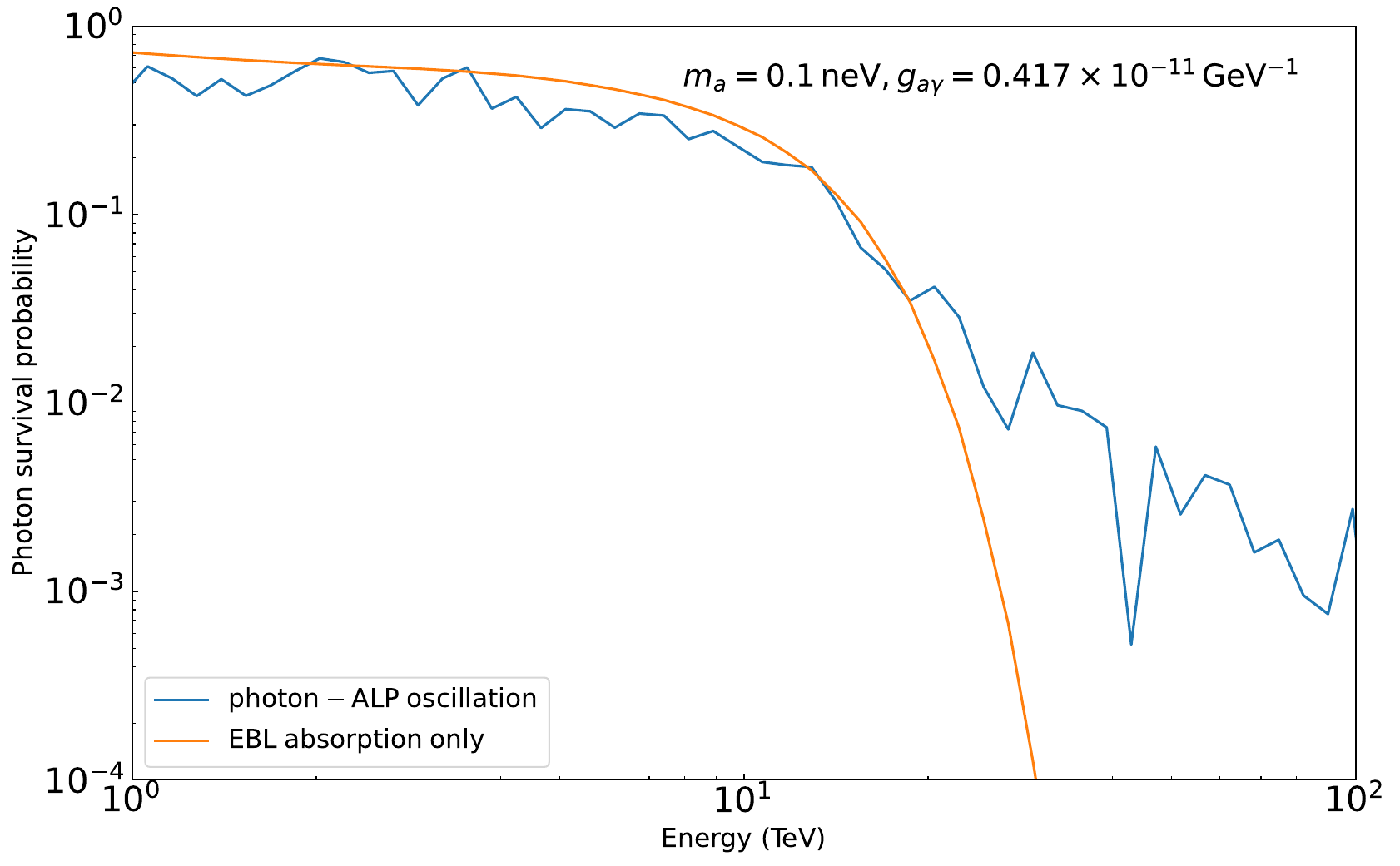}
\caption{The best-fit $\gamma$-ray SEDs of the Mrk 501 (left panels) and corresponding the photon survival probability $P_{\gamma\gamma}$ (right panels) in the above 1 $TeV$ energy range after propagation from the jet with helical and tangled magnetic field and galaxy cluster, where photon-ALP beams are produced, up to the Milky Way with Jansson $\&$ Farrar model by taking $g_{a\gamma\gamma}=0.417\times10^{-11}\mathrm~{GeV}^{-1}$ and $\mathrm{m_a}=10^{-10} eV$, where the goodness of fit is measured by employing the chi-square on every degree of freedom ($\chi^{2}$/d.o.f.) and residual plot. In the first row, the galaxy cluster with Gaussian turbulence magnetic field is postulated. In the second row, we consider the structured cavity magnetic field. {\it Left panels}: the dashed green line, dashed blue line and solid red line represent the intrinsic spectrum, the EBL absorption and the photon-ALP oscillation, respectively. The gray area indicates the $1\sigma$ confidence ban. The sensitivity for CTA, LHAASO 1-year and 5-year limits are represented by black dotted lines, black dashed lines, and black solid lines, respectively. {\it Right panels}: the orange solid line corresponds to conventional physics, and the blue solid line corresponds to photon-ALP oscillations.}
\label{sub:fig1}
\end{figure*}

\begin{figure*}
\centering
 \includegraphics[height=6cm,width=8.8cm]{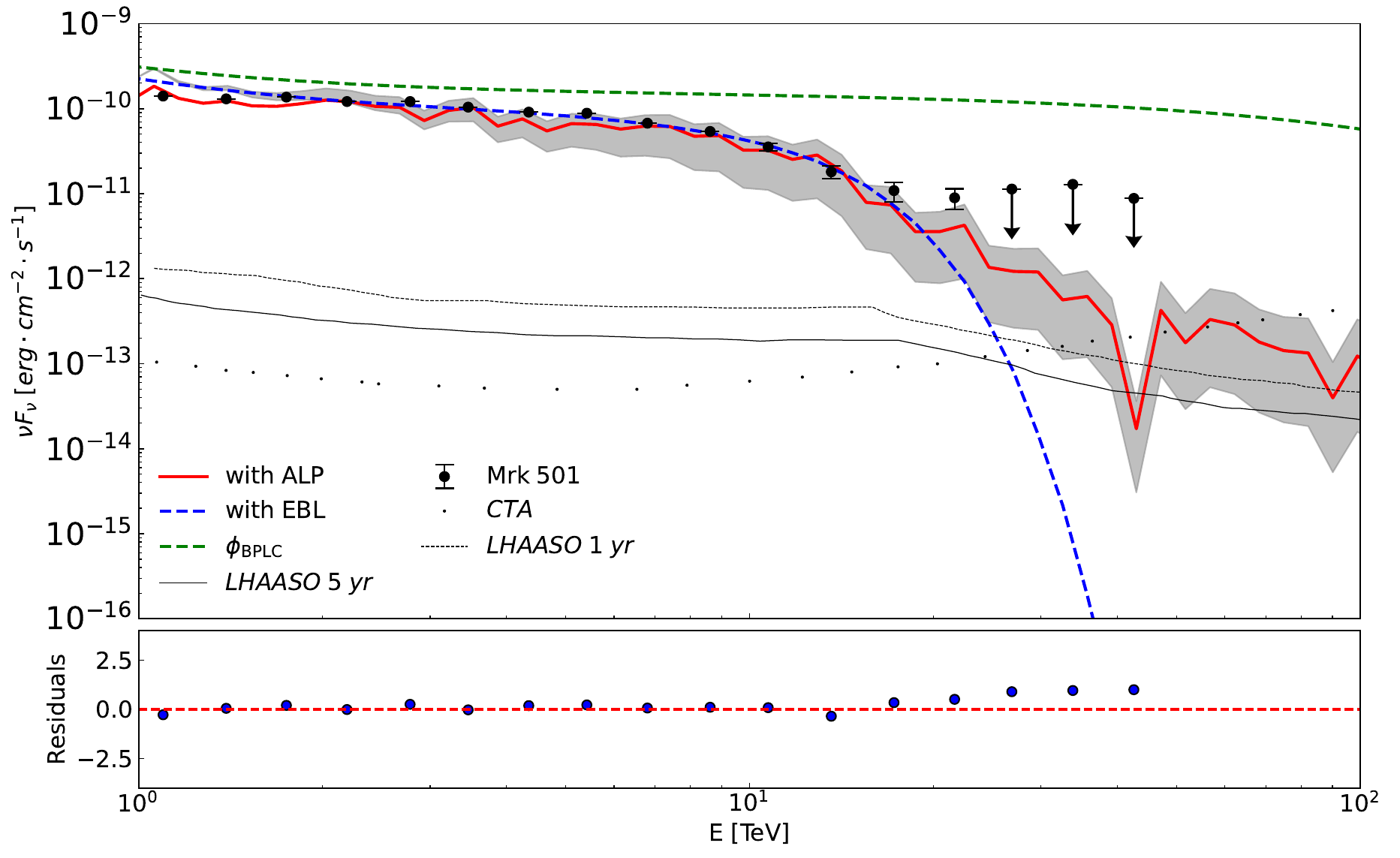}
 \includegraphics[height=6cm,width=8.8cm]{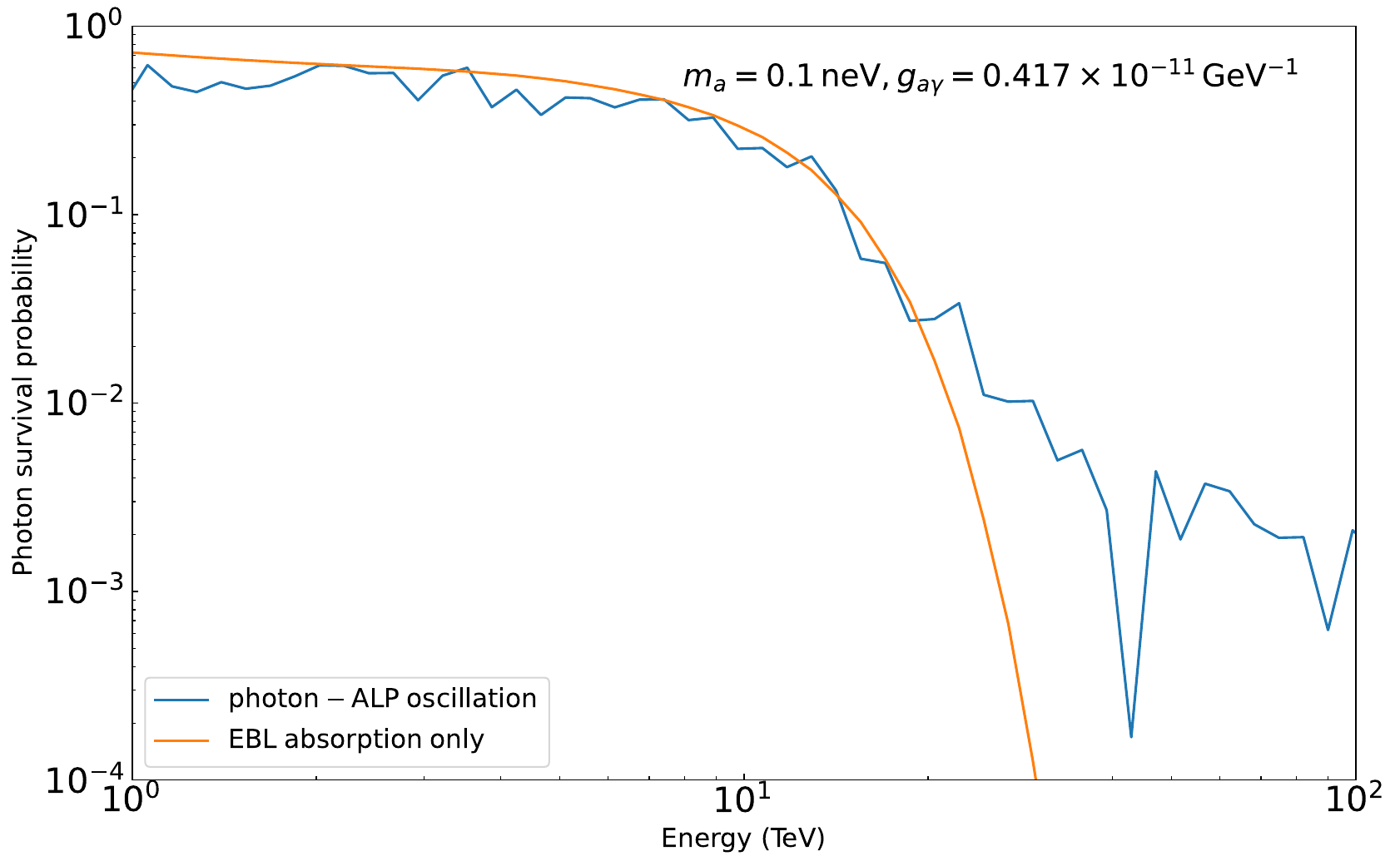}
 \includegraphics[height=6cm,width=8.8cm]{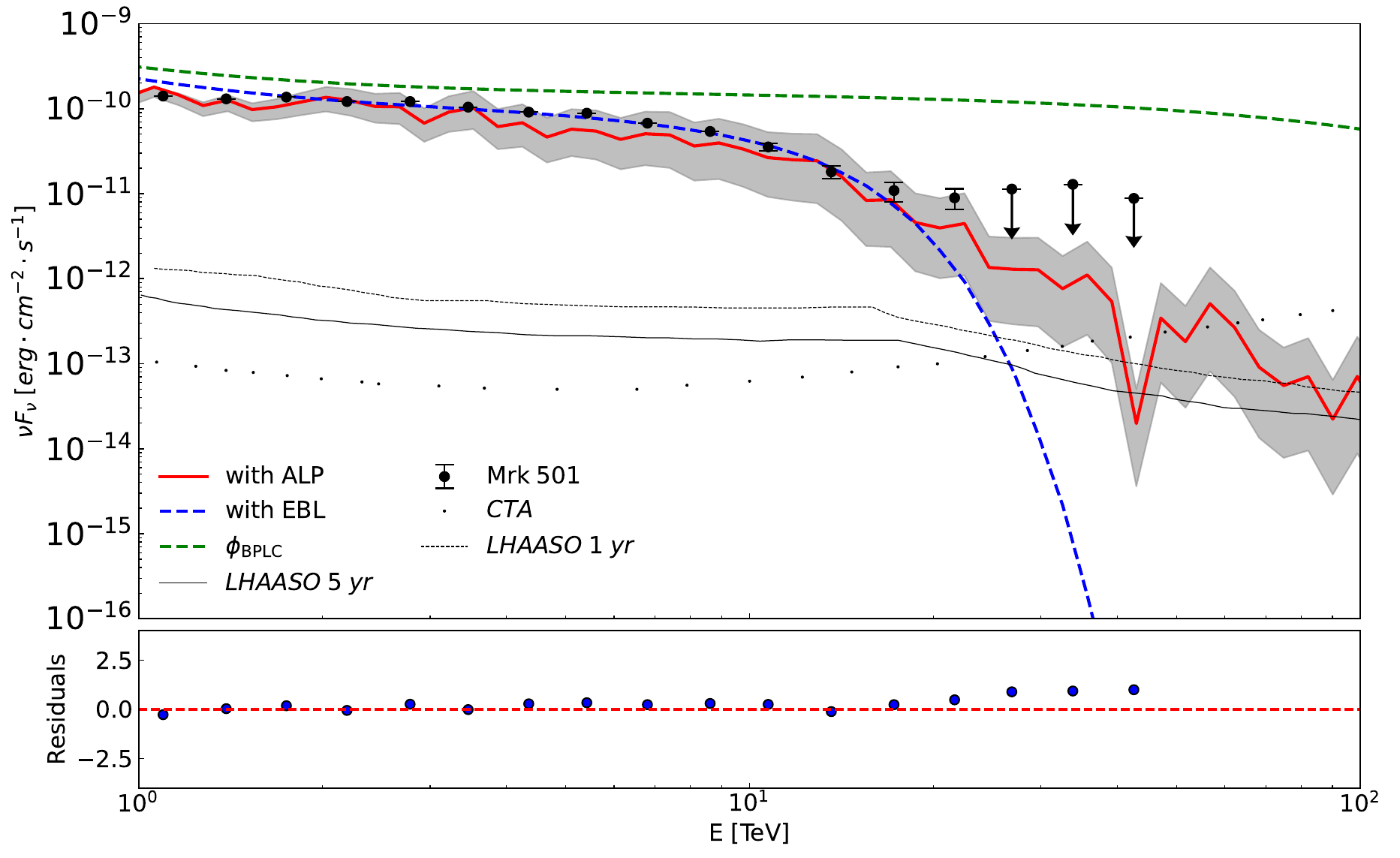}
 \includegraphics[height=6cm,width=8.8cm]{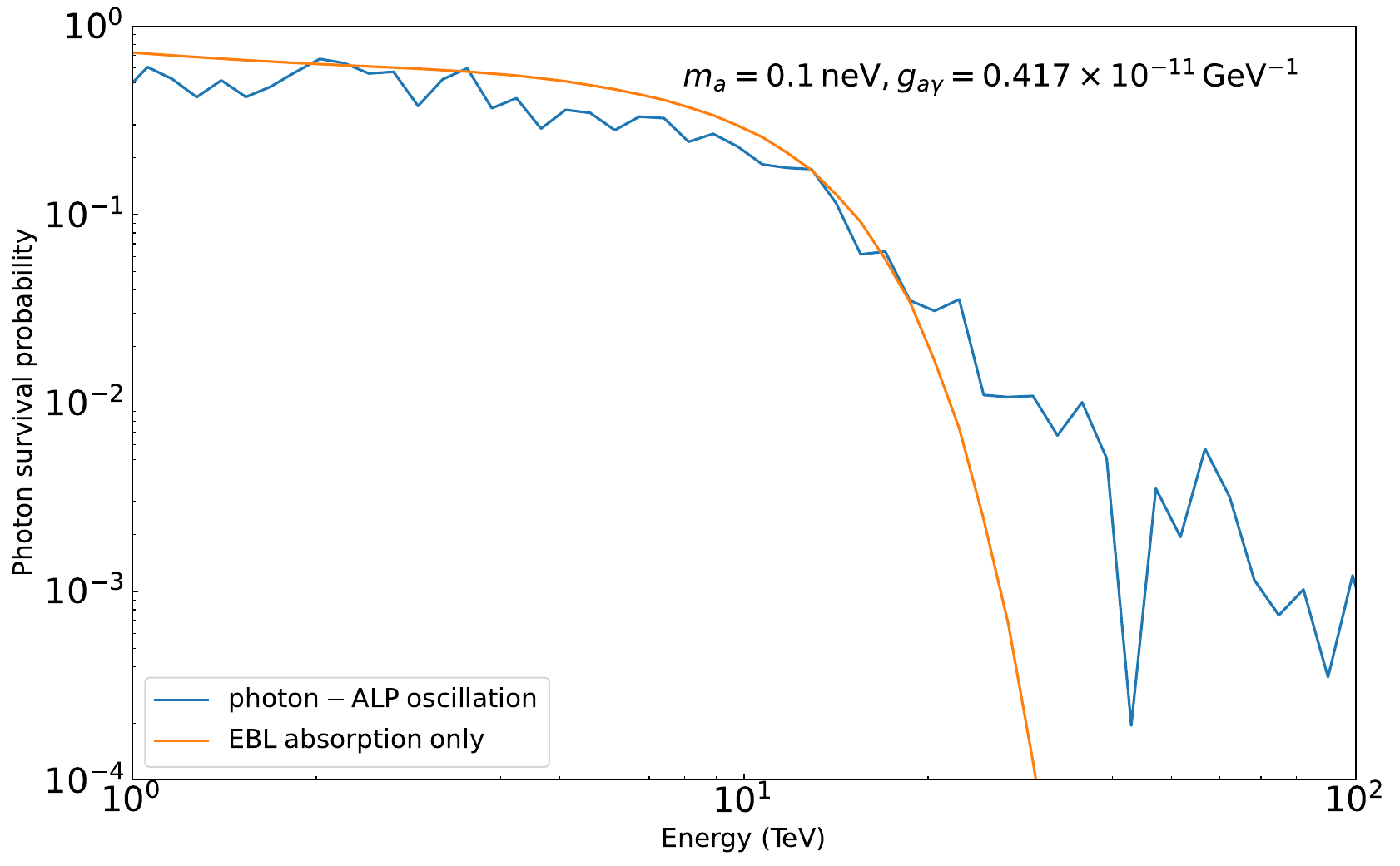}
\caption{Same as Fig.\ref{sub:fig1} but under the Pshirkov model of the Milky Way.}
\label{sub:fig2}
\end{figure*}

\begin{figure*}
\centering
 \includegraphics[height=6cm,width=8.8cm]{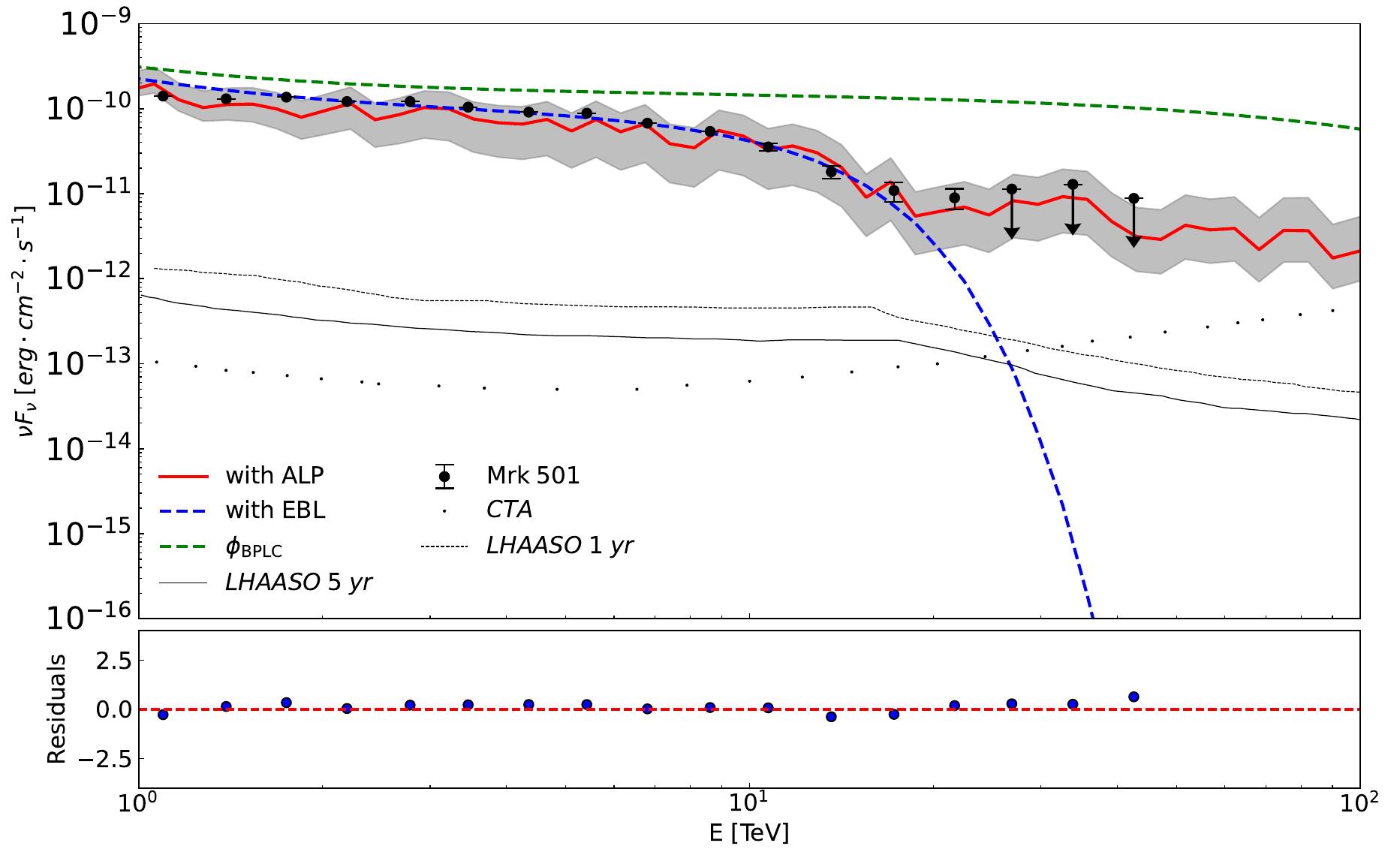}
 \includegraphics[height=6cm,width=8.8cm]{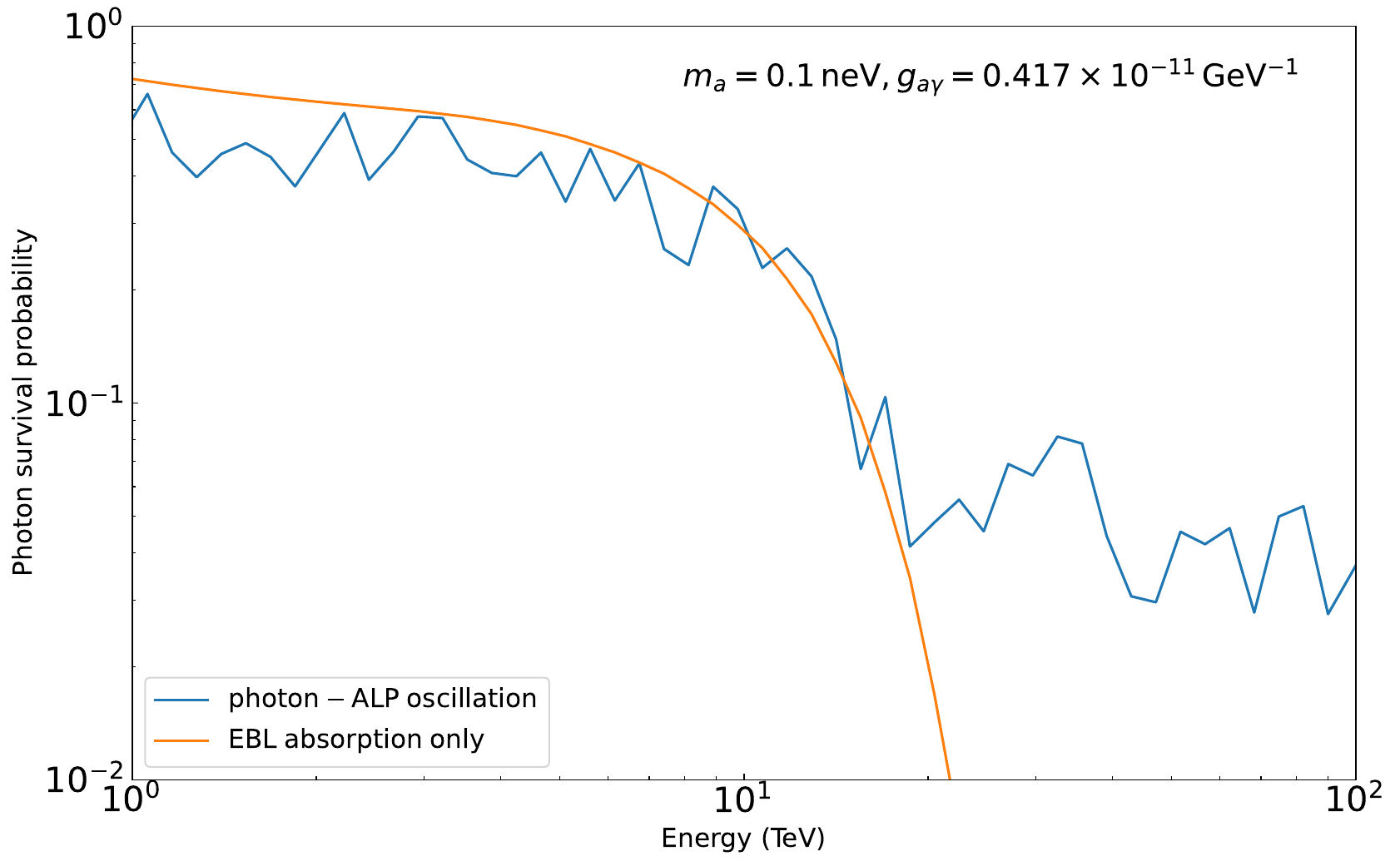}
 \includegraphics[height=6cm,width=8.8cm]{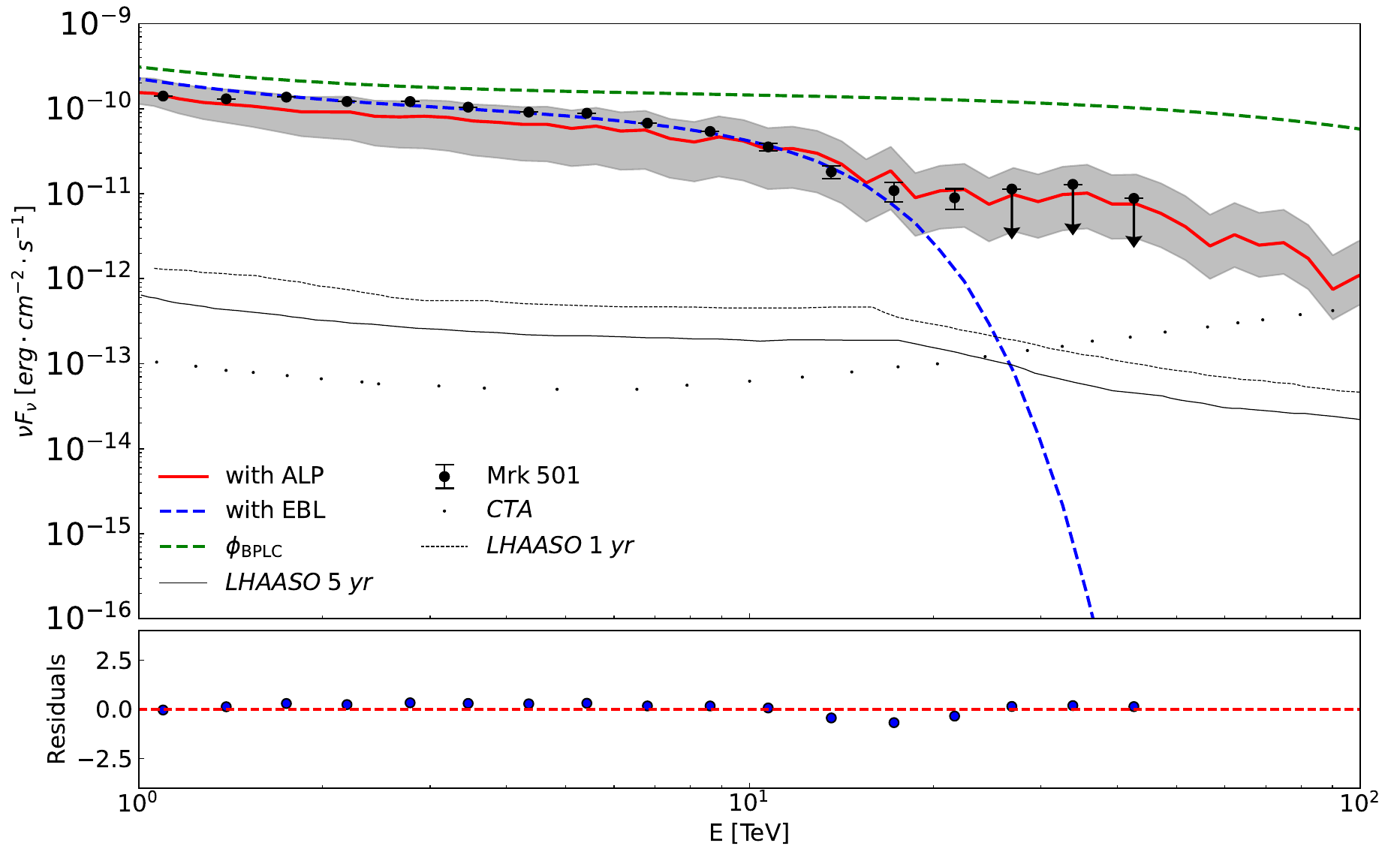}
 \includegraphics[height=6cm,width=8.8cm]{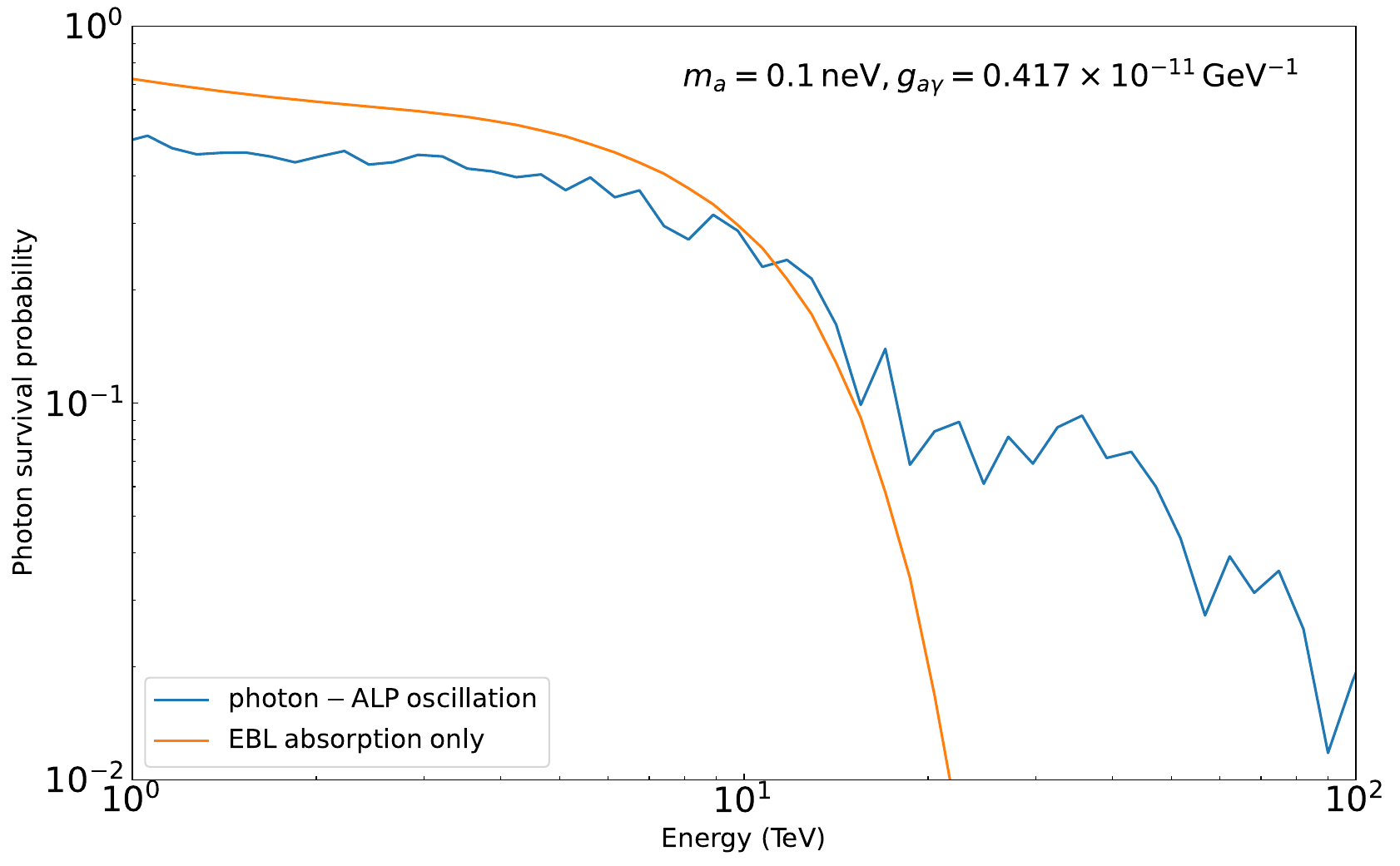}
\caption{ The photon-ALP beam are produce from the jet with simple toroidal magnetic field and galaxy cluster, up to the Milky Way with Jansson $\&$ Farrar model by taking $g_{a\gamma\gamma}=0.417\times10^{-11}\mathrm~{GeV}^{-1}$ and $\mathrm{m_a}=10^{-10} eV$. {\it Left panels}: the best-fit $\gamma$-ray SEDs of the Mrk 501 with the above 1 $TeV$ energy; {\it Right panels}: the photon survival probability $P_{\gamma\gamma}$ with the above 1 $TeV$ energy. In the first row, the galaxy cluster with Gaussian turbulence magnetic field is postulated. In the second row, we consider the structured cavity magnetic field.}
\label{sub:fig3}
\end{figure*}

\begin{figure*}
\centering
 \includegraphics[height=6cm,width=8.8cm]{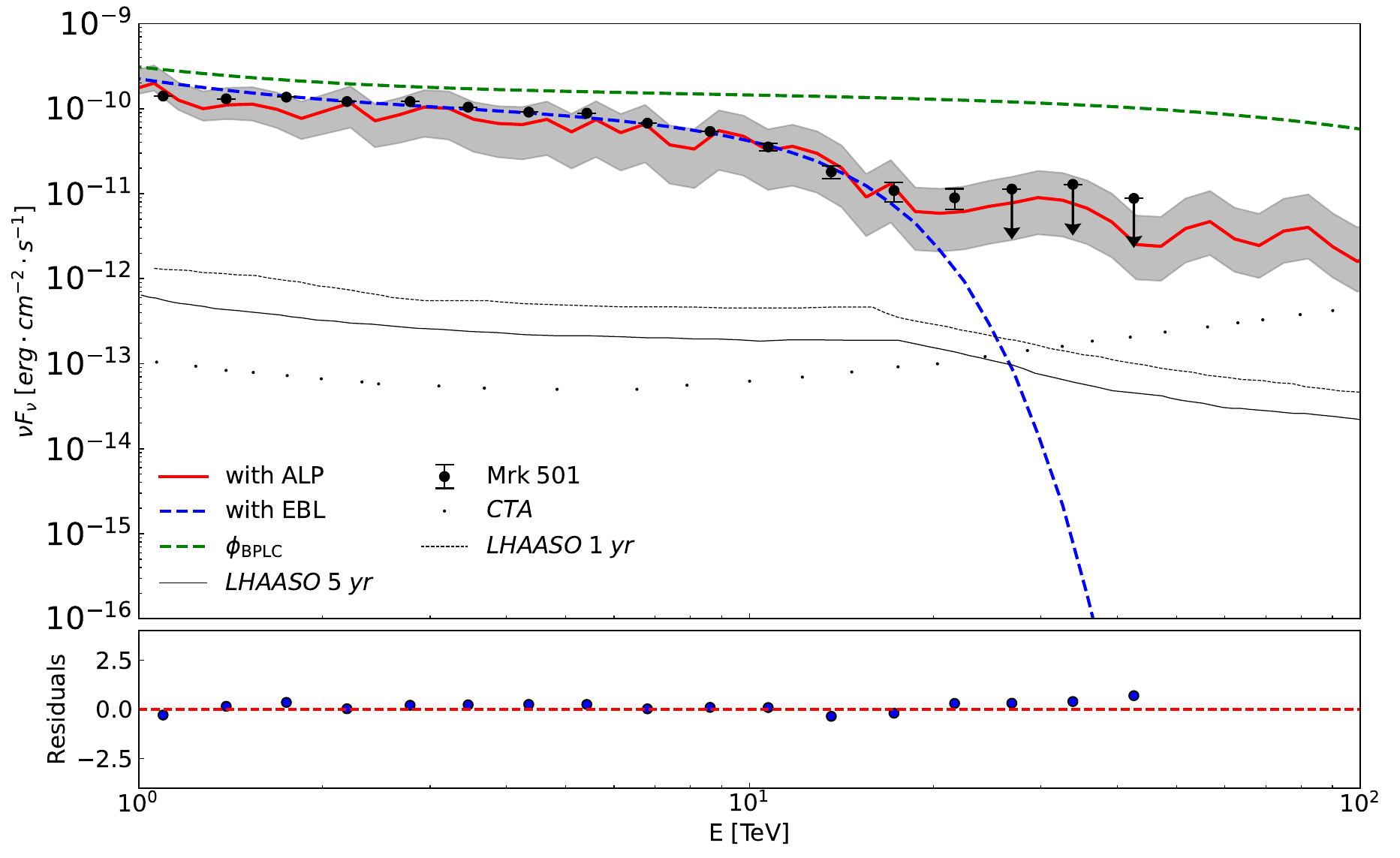}
 \includegraphics[height=6cm,width=8.8cm]{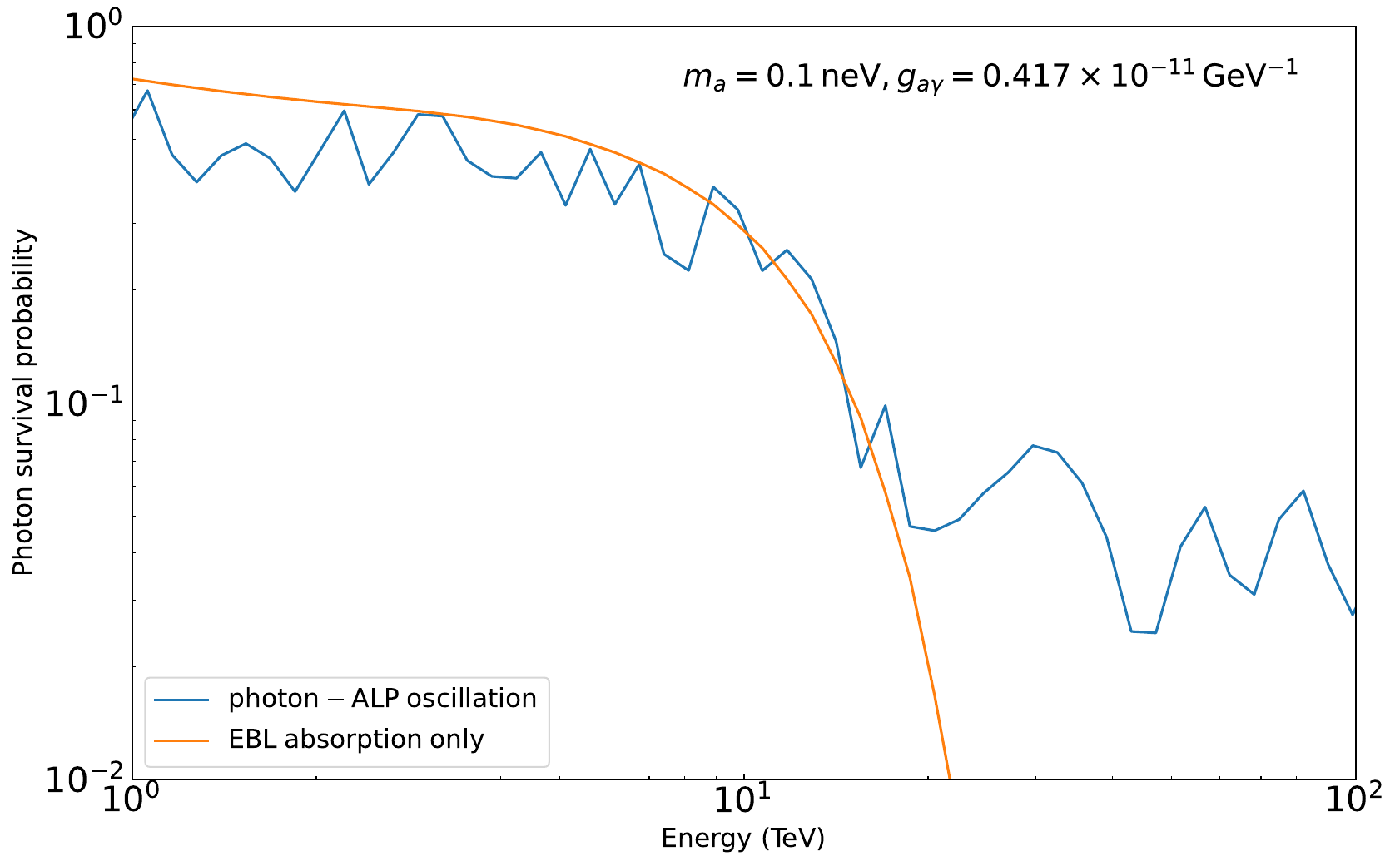}
 \includegraphics[height=6cm,width=8.8cm]{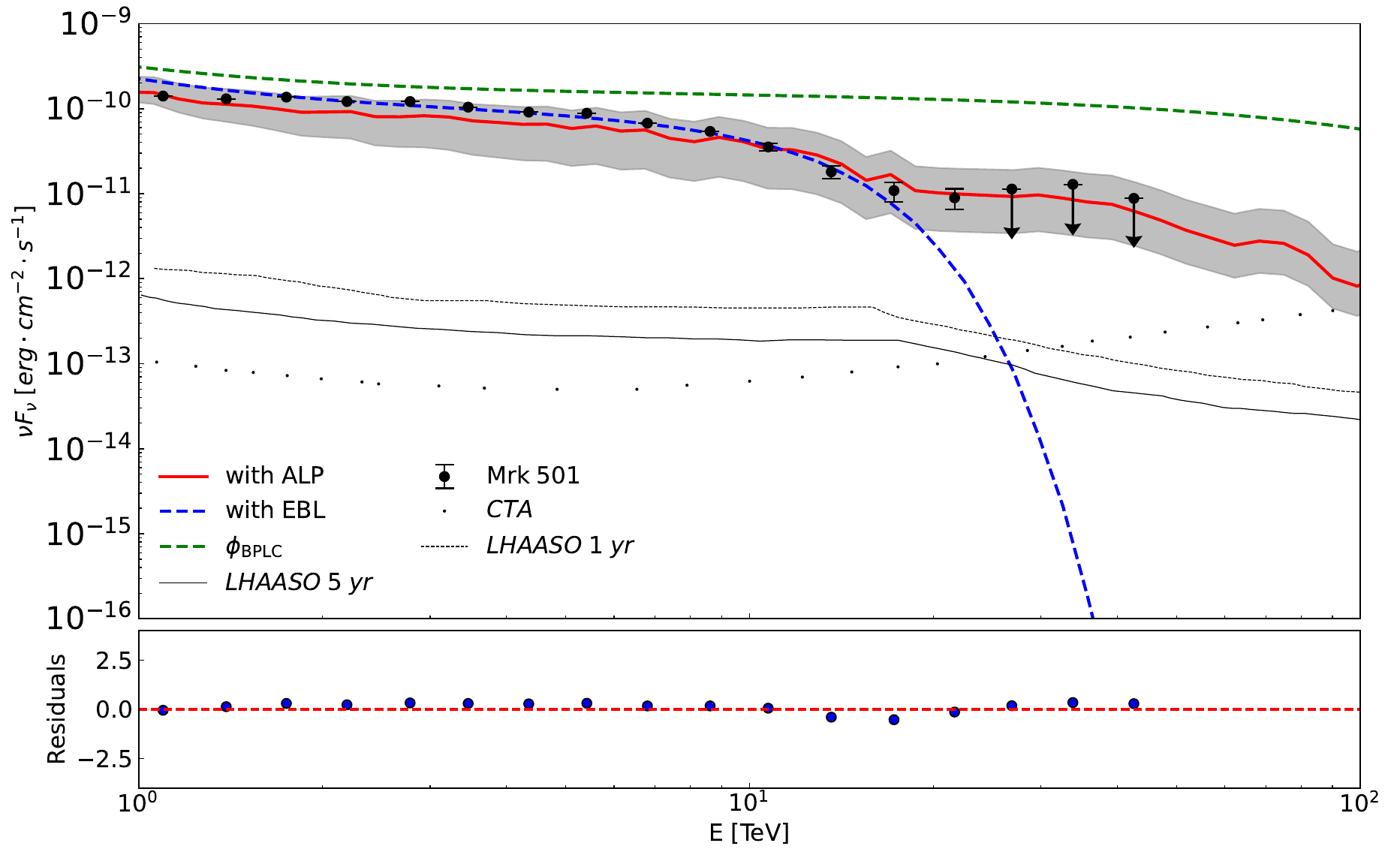}
 \includegraphics[height=6cm,width=8.8cm]{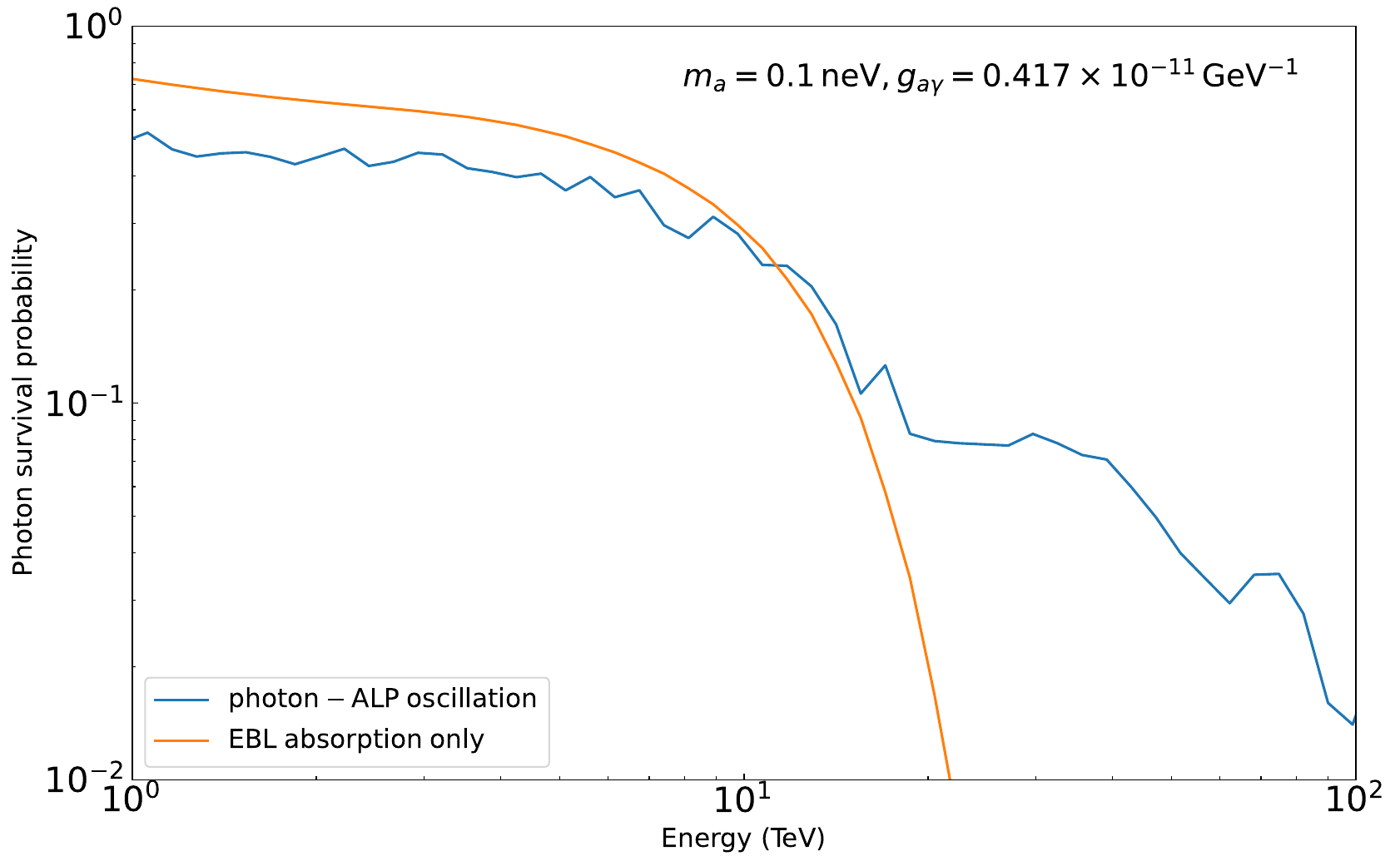}
\caption{Same as Fig.\ref{sub:fig3} but under the Pshirkov model of the Milky Way.}
\label{sub:fig4}
\end{figure*}

\subsection{Propagation in the jet} \label{sec:Propagation in the jet}

Blazars are a distinct class of active galactic nuclei (AGN) whose jets contain a great deal of important information about physical processes. Given the following three quantities: (1) the emission region distance from the central supermassive black hole (SMBH), denoted as $r_{\mathrm{em}}$, which is determined from measurements of the jet cross section, yielding $r_{\mathrm{em}}=10^{16}{-}10^{17}\mathrm~{cm}$ (for clarity, our fiducial choice is $r_{\mathrm{em}} = 3\times10^{16} \mathrm{cm}$) \citep{2023PhRvD.107d3006G,10.1093/mnras/stz1144}; (2) the transverse magnetic field profile $B_\mathrm{jet}(r)$; and (3) the electron density profile $n_\text{jet}(r)$. Consequently, we can evaluate the propagation of the photon-ALP beam within the jet, which terminates at a distance of approximately 1 kpc and enters the host galaxy.

In the jet region of BL Lac objects, the magnetic field can be described as follows: the poloidal component along the jet axis is represented as $B(r)\propto r^{-2}$, while the toroidal component transverse to the jet axis is represented as $B(r)\propto r^{-1}$.
The region of VHE photons radiating from the interior of the BL Lac jet is rather far from the central black hole \citep{2015PhLB..744..375T}. Therefore, we consider only the toroidal component of the magnetic field, which is transverse to the jet axis, while neglecting the poloidal component. The magnetic field profile can be implemented as \citep{1984RvMP...56..255B,2012SSRv..169...27P,10.1093/mnras/stt2133,1981ApJ...243..700K}
\begin{equation}
B_\mathrm{jet}(r)=B_{(0,jet)}\left(\frac r{r_{em}}\right)^{\eta_{\mathrm{jet}}},
\label{eq:7}
\end{equation}
with $\eta_{\mathrm{jet}}=-1$. Thanks to the conical shape of the jet, the electron number density $n_e^{\mathrm{jet}}$ profile can be represented by
\begin{equation}
n_\text{jet}(r)=n_{(0,jet)}\left(\frac r{r_{em}}\right)^{\xi_{\mathrm{jet}}},
\label{eq:8}
\end{equation}
with $\xi_{\mathrm{jet}}=-2$. For definiteness, we take the average values of the electron density $n_{(0,jet)}=5\times10^{4} \mathrm{cm}^{-3}$ \citep{2023PhRvD.107d3006G,10.1093/mnras/stz1144}. Becuase we compute the propagation of the photon-ALP beam crossing the jet in its comoving frame. Therefore, we apply the $E \to \Gamma E$ transformation to the beams to transition into the fixed frame of the subsequent regions, where $\Gamma$ denotes the Lorentz factor.

\subsubsection{Propagation in the jet-Mixing in helical and tangled component}
The Blandford-Znajek mechanism represents an important model for the launch of relativistic jets. In this process, the helical magnetic field lines originating from the accretion flow are wound up by the spinning black hole, thus extracting rotational energy from the central black hole and generating magnetically dominated jets at the poles.
\citep{10.1093/mnras/179.3.433,2018Galax...7....5G}. The particles within the jet can be accelerated by the magnetic energy until they approximately reach a state of equipartition.

Many rotation measure maps reveal asymmetries in the jet, indicating that the magnetic field is not solely transverse but indicative of the presence of large-scale helical magnetic fields. A helical magnetic field is expected to form with an initial pitch angle approximately determined by the ratio of the jet flow speed to the black hole rotation speed. The pitch angle increases along the jet as the jet expands and magnetic energy is dissipated to accelerate particles \citep{2018Galax...7....5G}. This helical magnetic field behavior is simulated by many relativistic jets \citep{10.1111/j.1745-3933.2009.00625.x, 1981ApJ...248...87L}. The entrainment of matter and the surrounding field would probably disrupt the magnetic field, leading to the emergence of a tangled component. A tangled field may influence the dominance of the transverse field, even at large radii, as a poloidal field with numerous reversals may still exhibit a net flux along the jet \citep{2021PhRvD.103b3008D}.

The jet magnetic field model is composed of a tangled component and a helical component, which transitions from poloidal to toroidal as it moves down the jet, as described in \cite{2021PhRvD.103b3008D}, where the helical component is described with two parameters and the tangled component with one parameter. The change rate of the transverse component of the helical field in the jet is determined by $\alpha$ ($B_{T}~\propto~r^{-\alpha}$), and the distance along the jet over which the helical component transitions to the toroidal component is governed by $r_{T}$. The helical field model was modified by introducing a tangled magnetic field component \citep{10.1093/mnras/133.1.67}. A constant fraction of the total magnetic energy density is the tangled component, as in \cite{10.1093/mnras/sts561} is expressed as
\begin{equation}
\frac{\langle B_{tangled}^{2}\rangle}{\langle B_{helical}^{2}\rangle}=\frac{f}{1-f},
\end{equation}
where $f$ represents the degree of entanglement. Increasing $f$ in the field reduces both the degree of asymmetry in the total intensity profiles and the degree of polarization. The impact of the tangled jet component $f$ on the spectrum is similar to that of the random domain in the intergalactic magnetic field, leading to small oscillations of the spectrum around its large-scale structure. The fiducial parameter values are derived from \cite{2021PhRvD.103b3008D}.

\subsubsection{Propagation in the jet-Mixing in simple toroidal magnetic field}

Assuming equipartition between the magnetic field and particle energies, the jet model postulates that the magnetic field strength follows the power-law given by Eq.(\ref{eq:7}) and the particle density follows the power-law given by Eq.(\ref{eq:8}). More details are provided in \ref{sec:Propagation in the jet} and Table \ref{Tab1}.

\subsection{Propagation in the galaxy cluster}

Observations show that Mrk 501 is located within a
small cluster \cite{1995ApJ...451...24S,1999bmtm.proc...57N}. Additionally, the photon-ALP mixing within the magnetic fields of galaxy clusters associated with Mrk 501 was studied in \cite{2012PhRvD..86g5024H}.

Faraday rotation measurements and non-thermal (synchrotron) emission at radio frequencies identify the presence of magnetic fields with the strengths of the order
of $\mathcal{O}(\mu\mathrm{G})$ in galaxy clusters.
The magnetic field distribution model $B_{\mathrm{ICMF}}\left(r\right)$ can be represented as
\begin{equation}
B_{\mathrm{ICMF}}(r)=B_{(0,\mathrm{ICMF})}\left(\frac{n_{\mathrm{ICMF}}(r)}{n_{(0,\mathrm{ICMF})}}\right)^{\eta_{\mathrm{ICMF}}},
\end{equation}
with $\eta_{\mathrm{ICMF}}=0.75$ \cite{2023PhRvD.107d3006G}.  Cold-core (CC) galaxy clusters are a class of galaxy clusters that are usually characterized by hosting an active galactic nucleus (AGN), with magnetic field strengths reaching up to tens of $\mu\mathrm{G}$ at the center of cooling core clusters \cite{2004IJMPD..13.1549G, 2012A&ARv..20...54F, 2013PhRvD..87c5027M, 2023PhRvD.107d3006G}, and the electron density distribution model $n_{\mathrm{ICMF}}(r)$ can be described by
\begin{equation}
n_{\mathrm{ICMF}}(r)=n_{(0,\mathrm{ICMF})}\left(1+\frac{r^{2}}{R_{core}^{2}}\right)^{{-\frac{3}{2}\xi_\mathrm{ICMF}}},
\end{equation}
where typically $\xi_\mathrm{ICMF} = 2/3$ , $n_{(0,\mathrm{ICMF})}=5\times10^{-2}\mathrm{cm}^{-3}$ \cite{2023PhRvD.107d3006G}, extension of the cluster ${r_{abell}} = 300 \mathrm{kpc}$ and the cluster core radius ${R_{core}} = 100 \mathrm{kpc}$
\citep{2010A&A...513A..30B,2011AA...529A..13K,Meyer_2014,2023PhRvD.107d3006G}.

\subsubsection{Propagation in the galaxy clusters-Mixing in Gaussian turbulence field}

In the case of photon-ALP oscillations, while early models described $B_{\mathrm{ICMF}}$ using a simple cell-like morphology, but a state-of-the-art characterization of $B_{\mathrm{ICMF}}$ is nowadays established \citep{Meyer_2014,2012PhRvD..86d3005W,2016PhRvL.116p1101A}.

$B_{\mathrm{ICMF}}$ exhibits isotropic Gaussian turbulence properties with a Kolmogorov-type power spectrum $M(k)\propto k^{q}$, where $k$ is the wavenumber in the range $k_{L}\leqslant k\leqslant k_{H}$ and zero otherwise. We take ${k_{L}} = 0.18 \mathrm{kpc}^{-1}$, ${k_{H}} = 3.14 \mathrm{kpc}^{-1}$ and the index $q=-11/3$ \cite{Meyer_2014,2008MNRAS.391..521L}. $B_{\mathrm{ICMF}}$ is expressed as
\begin{equation}
B_{\mathrm{ICMF}}(r)=\mathcal{B}\left(B_{(0,\mathrm{ICMF})},k,q,r\right)\left(\frac{n_{\mathrm{ICMF}}(r)}{n_{(0,\mathrm{ICMF})}}\right)^{\eta_{\mathrm{ICMF}}},
\end{equation}
where $\mathcal{B}$ denotes the spectral function characterizing the Kolmogorov-type turbulence in the cluster magnetic field. An analytical model of the turbulent and coherent magnetic fields in such a galaxy cluster was presented in \cite{Meyer_2014}; see also the assumed B-field environments in \cite{2016PhRvL.116p1101A,2011A&A...529A..13K} for more details.

\subsubsection{Propagation in the galaxy clusters-Mixing in structured cavity field}

X-ray and radio observations of the intracluster medium prove the existence of the cavity created by blazar jets. The cavity is filled with magnetized plasma from the jet and is observed through X-ray surface brightness depressions and intrinsic synchrotron emission. Stability demands that both toroidal and poloidal magnetic fields be present, with realistic configurations exhibiting vanishing magnetic fields at the boundary. In the embedding of unmagnetized plasma within an axisymmetric structure, the continuity of poloidal and toroidal magnetic field components at the bubble surface necessitates satisfying both the Dirichlet and Neumann boundary conditions of the elliptic Grad-Shafranov equation \cite{8224685}.

By assuming that gravity is negligible and focusing on magnetic cavities rather than self-gravitating systems such as magnetic stars, the equilibrium conditions are governed by 
\begin{equation}
J\times B=\nabla p,
\label{eq:(13)}
\end{equation}
where $B=\frac{\nabla P\times\mathbf{\hat{e}}_{\phi}+2I\mathbf{\hat{e}}_{\phi}}{r\sin\theta}$ is represented as axially symmetric magnetic fields. They set the speed of light to unity, and the force balance (\ref{eq:(13)}) gives the Grad-Shafranov equation
\begin{equation}
\Delta^{*}P+F(P)r^{2}\sin^{2}\theta+G(P)=0,
\end{equation}
where $\Delta^{*}=\frac{\partial^{2}}{\partial r^{2}}+\frac{\sin\theta}{r^{2}}\frac{\partial}{\partial\theta}\left(\frac{1}{\sin\theta}\frac{\partial}{\partial\theta}\right)$ is the Grad-Shafranov operator is due to the poloidal field force, $G(P)=4II^{\prime} (I=I(P))$ is due to the toroidal field force and $F(P)=4\pi\mathrm{d}p/\mathrm{d}P$ is due to the pressure gradient. The detailed process is taken from \cite{2004IJMPD..13.1549G, 8224685}.

\subsection{Propagation in the extragalactic space}

The morphology and strength of the extragalactic magnetic field $\mathbf{B}_{\mathrm{ext}}$ is known with difficulty. However, several configurations of $\mathbf{B}_{\mathrm{ext}}$ have been proposed \citep{1994RPPh...57..325K,GRASSO2001163,Wang_2016,2017PhRvD..96d3519M}.
It is generally assumed that $\mathbf{B}_{\mathrm{ext}}$ originate from quasars and primordial galactic outflows, and is modeled as a domain-like network \citep{1968Natur.217..326R,1969Natur.223..936H}. In this model,  $\mathbf{B}_{\mathrm{ext}}$ exhibits approximately uniform strength within each domain,  characterized by a coherence length $L_{\mathrm{dom}}$, while its direction changes randomly and discontinuously between adjacent domains \citep{1994RPPh...57..325K, GRASSO2001163}.
The coherence lengths range from $\text{1 Mpc}$ to $\text{10 Mpc}$, with $\mathbf{B}_{\mathrm{ext}}$ values falling within $10^{-7}~\mathrm{nG}\leq B_{\mathrm{ext}}\leq1.7~\mathrm{nG}$ \citep{2016PhRvL.116s1302P,2010Sci...328...73N}. But we have limited knowledge of the morphology and strength of $\mathbf{B}_{\mathrm{ext}}$. Therefore, we neglect the influence of $\mathbf{B}_{\mathrm{ext}}$ on the photon-ALP beam.

Another factor influencing $\gamma$-ray spectra in intergalactic space is the absorption by the extragalactic background light (EBL) \citep{2013APh....43..112D,2013A&A...550A...4H}. It is the cosmic electromagnetic background with wavelengths ranging from 0.1 to 1000 $\mu\mathrm{m}$ located in the ultraviolet to infrared wavelength range. When high-energy photons emitted from the source propagate through cosmic space, $\gamma$ ray photons may be absorbed by the EBL through the photon-photon pair production process \citep{2014JETPL.100..355R}. 
This causes an exponential attenuation of the $\gamma$-ray spectrum, quantified by the optical depth, which measures the dimming of the source \citep{2007ApJ...666..663B,2010ApJ...712..238F,2014JETPL.100..355R,2016RSOS....350555C}.  It is well known that photon-ALP oscillation effects can lead to a reduction of the cosmic opacity in a strong magnetic field and can substantially reduce the EBL absorption but not completely avoid it \citep{1983PhRvL..51.1415S,1988PhRvD..37.1237R}. Additionally, photon-ALP oscillations also explain the anomalous redshift dependence of blazar spectra compared to the effects of EBL absorption \citep{2020MNRAS.493.1553G,2023ApJ...952..152D}. We use the EBL model of \cite{brefId0, CrefId0}.

\subsection{Propagation in the Milky Way}

Faraday rotation measurements of the radio emission from the pulsar indicate that the strength of the Galactic magnetic field $\mathbf{B}_{\mathrm{MW}}$ is on the order of $\mathcal{O}(1)~\mu\mathrm{G}$. This field consists of both regular and turbulent components. We are focusing on the regular component of $\mathbf{B}_{\mathrm{MW}}$, which exerts a dominant effect on the propagation of the photon-ALP beam. Observations have shown that regular fields are composed of at least two parts: the ``disk" part is concentrated along the galactic plane and exhibits a roughly spiral shape with a typical scale height of about 1 $kpc$; the ``halo" part is located at some distance from the disk plane and usually regarded as a purely toroidal field with a vertical scale extending to above 3 $kpc$ \citep{2012PhRvD..86g5024H}.

We adopt the Jansson $\&$ Farrar model \citep{2012ApJ...757...14J, Jansson_2012}, which uses more rotation measures data and the latest WMAP7 synchrotron emission data. It reveals the existence of large-scale out-of-the-plane components of $\mathbf{B}_{\mathrm{MW}}$ and takes into account both striated fields and completely random fields. The latest data include polarized synchrotrons and different models of cosmic ray and thermal electron distributions, are detailed in \cite{2017ICRC...35..558U}.

We also consider an alternative Pshirkov model, which uses state-of-the-art Faraday rotation measurements of extragalactic radio sources to infer the $\mathbf{B}_{\mathrm{MW}}$ global structure \citep{2011ApJ...738..192P,2012PhRvD..86g5024H}. This model is based on both improved rotation measurement data and ionized gas distribution data. Furthermore, the numerical model has been refined to match these observations.

The electron number density in the Milky Way disk is $n_{e}^{\mathrm{MW}}\simeq1.1\times10^{-2}\mathrm{cm}^{-3}$, as inferred from the new free electron distribution model developed in \cite{2017ApJ...835...29Y}.

\section{Spectral analysis of blazar} \label{sec:Magnuss}

Once the intrinsic spectra are known, we use the overall photon survival probability to derive the corresponding observed spectra of Mrk 501 in the presence of photon-ALP oscillations from the BL Lac jet up to Earth. This approach allows us to compare the fitting results of the photon-ALP mixing model and the conventional physics model with the observational data.

As a first step, the observed and emitted spectra (number fluxes) can be described by
\begin{equation}
\phi(E)\equiv\frac{\mathrm{d}N}{\mathrm{d}t\mathrm{d}A\mathrm{d}E},
\end{equation}
where $\text{N}$ and $\mathrm{d}_{A}$ represent the VHE photon number and the infinitesimal area, respectively.

The EBL absorption-corrected spectra exhibit a broken power-law shape \citep{DrefId0}. Thus, we model the intrinsic spectrum using a broken power law with an exponential-cut-off (BPLC), which is described by \citep{2021PhRvD.104h3014L}
\begin{equation}
\phi_{\mathrm{BPLC}}=\phi_{0}(E/E_{0})^{-\alpha_{1}}[1+(\frac{E}{E0})^{g}]^{\frac{\alpha_{1}-\alpha_{2}}{g}}\exp(-E/E_{\mathrm{c}}),
\end{equation}
where $\phi_{0}$ is the flux normalization, $E_{0}=1000~\mathrm{GeV}$ is the fixed fiducial energy, the minimum cutoff energy $E_{\mathrm{c}}=100000~\mathrm{GeV}$, the photon index $a_{1}=1.6$, $a_{2}=0.6$, $g=1.6$ of the chosen intrinsic spectrum determined by the best fitting of $\Phi_{\mathbf{obs}}$ (Eq.(\ref{eq:17}), with EBL) to the data points.

The ALP-photon coupling constant $g_{a\gamma\gamma}=0.417\times10^{-11}~\mathrm{GeV}^{-1}$ and ALP mass $\mathrm{m_a}=10^{-10} eV$ were obtained from the previous work \citep{2023ApJ...952..152D} as benchmark values for the photon-ALP oscillation \citep{2022PhRvD.105j3034D}. $P_{\gamma\gamma}^{\mathrm{CP}}$ is due to EBL only in the case of conventional physics, while $P_{\gamma\gamma}^{\mathrm{ALP}}$ is due to EBL and photon-ALP oscillations in the case of the ALP model. The observed spectrum $\Phi_{\mathbf{obs}}(E,z)$ can be implemented as
\begin{equation}
\Phi_{\mathbf{obs}}(E,z)=P_{\gamma\gamma}\phi_{\mathrm{BPLC}},
\label{eq:17}
\end{equation}
where $P_{\gamma\gamma} = P_{\gamma\gamma}^{\rm CP}$ in the case of conventional physics and $P_{\gamma\gamma} = P_{\gamma\gamma}^{\rm ALP}$ in the case of the ALP model.

Moreover, the $\gamma$-ray SED is related to $\Phi_{\mathbf{obs}}(E,z)$ by
\begin{equation}
\nu F_{\nu}(\nu,E,z)=E^{2}\Phi_{\mathrm{obs}}(E,z),
\end{equation}
where $F_{\nu}(\nu,E,z)$ is the specific apparent luminosity.

We consider the reduced chi-square $\chi^{2}$/d.o.f. in the fitting for the observations of the High-Energy Gamma-Ray Astronomy (HEGRA). The corresponding $\chi^{2}$/d.o.f.-function is expected to be represented by
\begin{equation}
    \chi^{2}=\frac{1}{N-dof}\sum_{\mathrm{i}=1}^{\mathrm{N}}\frac{({\Phi}_{obs,i}-\Phi_{i})^{2}}{\delta\Phi_{obs,i}{}^{2}},  
\end{equation}
where $N$, $dof$, ${\Phi}_{obs,i}$, $\Phi_{i}$, and $\delta\Phi_{obs,i}{}^{2}$ represent the number of observed data points,  the degrees of freedom (the number of free parameters used for the model), the observed value, modeled value, and uncertainty of the observed photon flux in the i-th energy bin, respectively.  We take $20\%$ of the observed $\gamma\mathrm{-ray}$ flux as the error for the data points without error \citep{2021ApJ...915...59Z}. The free parameter values of the photon-ALP mixing model are shown in Table \ref{Tab1}.

\subsection{Application of the model to Mrk 501}

Mrk 501 is a bright high-frequency peaked BL Lacs with the redshift $z=0.034$ whose quiescent states have been studied as a possible environment for exploring new physical signals \citep{Fairbairn_2014}. The active state of Mrk 501 is certainly the best candidate for such an issue \citep{ErefId0}. In this fashion, we use observations from the HEGRA \citep{2001ApJ...546..898A}, where the high-energy emission of Mrk 501 may be above $10 \;\rm{TeV}$, which allows us to characterize its spectrum well, since at such high energies, it may produce different predictions compared with conventional physics.

We analyze the photon survival probability $P_{\gamma\gamma}^{\rm ALP}$ and the corresponding $\gamma$-ray SED resulting from the propagation of the photon-ALP beam crossing the several different magnetized media discussed in Sec.\ref{sec:freezeout} (blazar jet, galaxy cluster, extragalactic space, and Milky Way). We research the propagation of the photon-ALP beam by considering eight different scenarios: 1) The photon-ALP beam is generated in the jet with helical entangled magnetic fields and galaxy cluster with Gaussian turbulent magnetic fields, up to the Milky Way using the Jansson $\&$ Farrar model, where the $\chi^{2}$/d.o.f.=1.58 is obtained by employing Eq.(18) in order to consider the degrees of freedom. The left of the top panel in Fig.\ref{sub:fig1} shows the VHE $\gamma$-ray SED from the conventional physics model and the photon-ALP mixing model, respectively, compared with observed data at different energy values, where the conventional physics mode with the $\chi^{2}$/d.o.f.=3.02 is obtained by employing Eq.(18). And the small panel below the top left figure shows the residual plot between the VHE $\gamma$-ray SEDs with photon-ALP oscillations and the observations. The blue dashed curve, the red solid curve and the green dashed curve represent conventional physics, photon-ALP oscillations and intrinsic spectrum, respectively. However, the black dotted line, the black dashed line and the black solid line show the sensitivity limits of CTA, 1-year and 5-year for LHAASO respectively. The right of the top panel of Fig.\ref{sub:fig1} shows the behavior of $P_{\gamma\gamma}$ versus the observed energy following the same procedure as the photon-ALP beam propagation above. The orange solid curve corresponds to conventional physics, and the blue solid curve represents photon-ALP oscillations. 2) The bottom panel of Fig.\ref{sub:fig1} shows the same case as the top panel, but the magnetic field structure of galaxy cluster is the structured cavity magnetic field, where the $\chi^{2}$/d.o.f.=2.37 is obtained by employing Eq.(18). 3) The top panel of Fig.\ref{sub:fig2} show the photon-ALP beam generated in the jet with helical entangled magnetic fields and galaxy cluster with Gaussian turbulent magnetic fields, up to the Milky Way using the Pshirkov model, where the $\chi^{2}$/d.o.f.=1.52 is obtained by employing Eq.(18). 4) The bottom panel of Fig.\ref{sub:fig2} shows the same case as the top panel, but the magnetized structure of the galaxy cluster is the structured cavity magnetic field, where the $\chi^{2}$/d.o.f.=2.45 is obtained by employing Eq.(18). 5) The top panel of Fig.\ref{sub:fig3} shows the photon-ALP beam generated in the jet with simple toroidal magnetic fields and galaxy cluster with Gaussian turbulent magnetic fields, up to the Milky Way using the Jansson $\&$ Farrar model, where the $\chi^{2}$/d.o.f.=1.42 is obtained by employing Eq.(18). 6) The bottom panel of Fig.\ref{sub:fig3} shows the same case as the top panel, but the magnetic field structure of the galaxy cluster is the structured cavity magnetic field, where the $\chi^{2}$/d.o.f.=2.25 is obtained by employing Eq.(18). 7) The top panel of Fig.\ref{sub:fig4} shows the photon-ALP beam generated in the jet with simple toroidal magnetic fields and galaxy cluster with Gaussian turbulent magnetic fields, up to the Milky Way using the Pshirkov model, where the  $\chi^{2}$/d.o.f.=1.53 is obtained by employing Eq.(18). 8) The bottom panel of Fig.\ref{sub:fig4} shows the same case as the top panel, but the magnetic field structure of the galaxy cluster is the structured cavity magnetic field, where the $\chi^{2}$/d.o.f.=2.01 is obtained by employing Eq.(18). The concomitant behavior of energy oscillations in the spectrum would provide conclusive evidence of photon-ALP oscillations. However, the different magnetic field scenarios obviously impact the VHE $\gamma$-ray SEDs with photon-ALP oscillations.

\begin{table}{}
\centering
\fontsize{8}{11}\selectfont
\caption{Physical free parameters of the model spectra}\label{Tab1}
\begin{tabular}{ccccc}
\hline\hline
\multicolumn{2}{c}{{jet with helical and tangled component}}& {} &\multicolumn{2}{c}{{cluster magnetic field}} \\
\cline{1-2}\cline{4-5}
Parameters & Value & & Parameters & Value \\
\hline
${\Gamma}$ & $\text{8}$ & & $B_{(0,\mathrm{ICMF})}$ & $28~\mu\mathrm{G}$\\
$B_{(0,\mathrm{jet})}$ & $\text{0.1~G}$ & & - & -  \\

\hline
\end{tabular}\\
\end{table}

\begin{table}{}
\centering
\fontsize{8}{11}\selectfont
\begin{tabular}{ccccc}
\hline\hline
\multicolumn{2}{c}{{jet with simple toroidal magnetic field}}& {} &\multicolumn{2}{c}{{cluster magnetic field}} \\
\cline{1-2}\cline{4-5}
Parameters & Value & & Parameters & Value \\
\hline
${\Gamma}$ & $\text{6}$ & & $B_{(0,\mathrm{ICMF})}$ & $26~\mu\mathrm{G}$\\
$B_{(0,\mathrm{jet})}$ & $\text{0.5~G}$ & & - & -  \\

\hline
\end{tabular}\\
\end{table}

\section{Discussion and Conclusion} \label{sec:Conclusion and Discussion}

This paper discusses the impact of photon-ALP oscillations on $\gamma$-ray spectra from distant sources in astrophysical magnetic fields.

We find that 1) when the photon-ALP beam is produced from the jet with helical and tangled magnetic fields. Based on the calculated reduced chi-square value, and in scenarios 1) (see the left of the top panel of Fig.\ref{sub:fig1}) and scenarios 3) (see the left of the top panel of Fig.\ref{sub:fig2}), the $\gamma$-ray SED with photon-ALP oscillations for Mrk 501 naturally matches data points, especially in the sub-tens of TeV energy range. However, the $\gamma$-ray SED of scenarios 2) (see the left of the bottom panel of Fig.\ref{sub:fig1}) and scenarios 4) (see the left of the bottom panel of Fig.\ref{sub:fig2}) do not match the data points well. However, the photon survival probabilities in the right panels of Fig.\ref{sub:fig1} look very similar and the same is true for the photon survival probabilities in the right panels of Fig.\ref{sub:fig2}. This study shows that in the presence of the jet with helical and tangled magnetic fields, the propagation scenario does not significantly affect the total photon survival probability, whether the photon-ALP beam crosses the galaxy cluster with a Gaussian turbulence field or a cavity field.

2) When the photon-ALP beam is produced from the jet with simple toroidal magnetic fields. We analyze the calculated reduced chi-square value, and in scenarios 5) (see the left of the top panel of Fig.\ref{sub:fig3}) and scenarios 7) (see left of the top panel of Fig.\ref{sub:fig4}), the $\gamma$-ray SED with the photon-ALP oscillation for Mrk 501 is in better agreement with data points, especially towards the high energy range. However, the $\gamma$-ray SED of scenarios 6) (see the left of the bottom panel of Fig.\ref{sub:fig3}) and scenarios 8) (see the left of the bottom panel of Fig.\ref{sub:fig4}) do not match the data points well.

The photon-ALP oscillation effect on the $\gamma$-ray spectrum becomes particularly apparent at high energies, especially in the presence of ubiquitous astrophysical magnetic fields \citep{Majumdar_2018,2019PhRvD.100l3004X,2022PhRvD.106h3020E,2023PhRvD.107j3007G}. In particular, photon-ALP beams that cross different magnetic field structures retain information about all magnetic field structures they traverse, amplifying the variation in the oscillation amplitude. Our results show that (1) The photon-ALP oscillation model can fit the observational data at sufficiently high energies in a better way with respect to conventional physics, partially reducing EBL absorption. (2) By using the physical parameters from table \ref{Tab1}. In the case of the Gaussian turbulent magnetic field, the $\gamma$-ray flux is typically matched to the observational data compared with the structured cavity magnetic field, this is because the Gaussian turbulent magnetic field is a divergence-free, homogeneous, and isotropic field with zero mean and variance $\mathcal{B}^{2}$, which better describes the observed multi-scale fluctuations \cite{2004A&A...424..429M, Meyer_2014}. In the case of Jansson $\&$ Farrar model, the photon survival probability $P_{\gamma\gamma}^{\rm ALP}$ and the resulting $\gamma$-ray flux are generally larger compared to the Pshirkov model, especially in energy regions exceeding $10 \;\rm{TeV}$, this enhancement is attributed to the presence of a large vertical scale height in the halo magnetic field and the presence of a poloidal ``X-shaped" component extending out-of-plane from the Galactic center. This will benefit the detect the hardening VHE spectra from extragalactic sources \citep{2012PhRvD..86g5024H}. However, there is not much difference between the plots derived with the Janson $\boldsymbol{\&}$ Farrar model and those obtained from the Pshirkov one. The primary factor could be the dependence of the Milky Way's realistic magnetic field on the position of the source in the sky. Consequently, the back-conversion probability $P_{a\gamma}$ for ALPs to convert into photons in the Milky Way also depends on the source position. According to \citep{2012PhRvD..86g5024H}, a skymap illustrating the back-conversion probability in galactic coordinates for Mrk 501 has been provided, using both the Jansson $\boldsymbol{\&}$ Farrar or Pshirkov magnetic field models. We find that the $P_{a\gamma}$ for Mrk 501 exhibits no significant difference when these two distinct magnetic field models are compared.

In this paper, we mainly discuss the hardening of the TeV spectrum of Mrk 501 by the photon-ALP oscillation effect. In fact, the spectrum hardening may also be due to Lorentz invariance violation (LIV) effects  \cite{2010JMatR..25.1251P, 2023PhRvD.107h3001Z}, Bethe-Heitler (BH) cascade processes \cite{2010APh....33...81E, 2016A&A...585A...8Z}, and EBL
uncertainties \cite{2020MNRAS.491.5268G}. Distinguishing which scenario is at play requires constructing more refined physical models and higher spectral resolution measurements. Maybe the detailed spectral observations from the next generation of ultra-high energy $\gamma$-ray detectors with higher spectral resolution will provide us with opportunities.

\section*{Acknowledgements}

We thank the anonymous referee for constructive comments and suggestions. This work would like to thank this webpage \footnote{ https://gammaalps.readthedocs.io}. This work is partially supported by the National Natural Science Foundation of China (grant Nos. 12363002 and 12163002).

\bibliography{main}

\end{document}